\documentclass[11pt]{article}
\usepackage{eurosym}
\usepackage{amsfonts}
\usepackage{amssymb}
\usepackage{graphicx}
\usepackage{amsmath}
\usepackage{makeidx}
\usepackage{indentfirst}
\usepackage[T1]{fontenc}
\usepackage[utf8]{inputenc}

\newcounter{resultnum}[section]
\newcounter{conclusionnum}[section]
\newcounter{conditionnum}[section]
\newcounter{conjecturenum}[section]
\newcounter{examplenum}[section]
\newcounter{exercisenum}[section]
\newcounter{lemmanum}[section]
\newcounter{notationnum}[section]
\newcounter{theoremnum}[section]
\newcounter{definitionnum}[section]
\newcounter{corollarynum}[section]
\newcounter{remarknum}[section]
\newcounter{propositionnum}[section]
\newcounter{acknowledgementnum}[section]
\newcounter{algorithmnum}[section]
\newcounter{axiomnum}[section]
\newcounter{casenum}[section]
\newcounter{claimnum}[section]
\newcounter{summarynum}[section]
\newcounter{problemnum}[section]
\begin{document}

\title{Off-diagonal deformations of regular Schwarzschild black holes\\ and general relativistic G. Perelman thermodynamics}
\date{%Mar 26 - \ 
Apr 5, 2025}
\author{ \textbf{Sergiu I. Vacaru} \thanks{%
emails: sergiu.vacaru@fulbrightmail.org ; sergiu.vacaru@gmail.com }  \\   
{\small \textit{\ Department of Physics, Kocaeli University, Izmit, 41001, T%
\"{u}rkiye; }}  \\  {\small \textit{Department of Physics, California
State University at Fresno, Fresno, CA 93740, USA}} \and {\textbf{El\c{s}en
Veli Veliev }} \thanks{email: elsen@kocaeli.edu.tr and elsenveli@hotmail.com} \\
%EndAName
{\small \textit{\ Department of Physics,\ Kocaeli University, Izmit, 41001, T\"{u}rkiye }} \vspace{.1 in} }
\maketitle

\begin{abstract}
We construct new classes of solutions describing generic off-diagonal deformations of regular Schwarzschild black holes (BHs) in general relativity (GR). Examples of such (primary) diagonal metrics reducing the Einstein equations to integrable systems of nonlinear ordinary differential equations were studied in a recent work by R. Casadio, A. Kamenshchik and J. Ovalle in Phys. Rev. D 111 (2025) 064036. We develop and apply our anholonomic frame and connection deformations method, which allows us to generate new classes of target off-diagonal solutions. Ansatz that reduces the gravitational field equations to  systems of (exactly or parametric) integrable systems of nonlinear partial differential equations are used.  We find and analyze certain families of deformed regular BHs containing an off-diagonal de Sitter condensate encoding solitonic vacuum configurations, with possible deformations of horizons and/or gravitational polarizations of constants. We emphasize that general off-diagonal solutions do not involve certain hypersurface or holographic configurations. This means that the Bekenstein-Hawking thermodynamic paradigm is not applicable for characterizing the physical properties of such target regular solutions. We argue that the concept of G. Perelman's entropy and relativistic geometric flow thermodynamics is more appropriate. Using nonlinear symmetries involving effective cosmological constants, we show how to compute  thermodynamic variables for various classes of physically essential solutions in GR. 
%%%

\end{abstract}

\tableofcontents

% \newpage

%%%

\section{Introduction,  geometric preliminaries, and objectives}

\label{sec1}Recent years have witnessed a rise in interest in the study of regular solutions in general relativity (GR) and modified gravity theories (MGT) as an alternative to singular counterparts \cite{casadio25} (see also references therein). The main goal of this work is to elaborate on new geometric methods for generating off-diagonal regular solutions in GR \cite{sv11,vacaruplb16,vbubuianu17,partner02}. We formulate and analyze the conditions when generic off-diagonal gravitational interactions transform singular or regular variants of Schwarzschild black hole (BH) metrics into solutions embedded into nontrivial gravitational vacuum, with possible deformations of horizons, polarization of physical constants, pattern-forming structures, etc. The approach is elaborated in the framework of a relativistic generalization \cite{svnonh08,gheorghiuap16,partner06} of the G. Perelman thermodynamics \cite{perelman1}, which can be used for characterizing fundamental physical properties of off-diagonal solutions in GR and MGTs.
%%%%%%%

\vskip5pt The GR theory is formulated in standard form as a geometric model on a four-dimensional (4-d) Lorentz manifold $V.$ Such a curved spacetime is
enabled with a pseudo-Riemannian metric, $g,$ and a corresponding Levi-Civita (LC) connection, $\nabla =\nabla \lbrack g],$ structures. For details, we cite the monographs \cite{hawking73,misner73,wald82,kramer03}.\footnote{By definition, $\nabla $ is a linear connection which is metric compatible and with zero torsion. We consider that readers are skilled in mathematical relativity and differential geometry. In our approach, priority is given to an abstract geometric formulation of GR and fundamental equations when nonholonomic and holonomic frames, distortion of connections, respective indices etc. are standard ones from the geometry of relativistic manifolds.} The geometric objects $(g,\nabla )$ are subjected to the condition to define solutions of the Einstein equations, 
\begin{equation}
Ric[\nabla ]=\Upsilon \lbrack g,\nabla ].  \label{einststand}
\end{equation}%
This system of nonlinear partial differential equations (PDEs)  is written in abstract geometric form \cite{misner73} using the Ricci tensor, $Ric,$ and effective sources, $\Upsilon ,$ correspondingly defined by physically important energy-momentum-tensors of matter fields. A source $\Upsilon $ is proportional to the gravitational constant and includes a term $-\lambda g$ defined by a nontrivial cosmological constant $\lambda .$ We write $[\nabla],\nabla \lbrack g]),...,$ or $[g,\nabla ]$ to state respective functional, or coordinate, dependence of geometric and physical objects. Our system of notations is formulated for elaborating on new advanced geometric methods of constructing off-diagonal solutions in gravity theories, which is slightly different from the conventions used in \cite{hawking73,misner73,wald82}.
%%%%%%%%%%

The bulk of important solutions in GR (various types of black holes, BHs,  or homogeneous and isotropic cosmological metrics) were constructed for diagonal ansatz with coefficients depending, for instance, on a radial, or a time-like coordinate \cite{kramer03}. Typically, the solutions of (\ref{einststand}) are generated  for certain imposed spherical, cylindrical, and rotational symmetries. The integration constants for such solutions are defined by some asymptotic/ boundary conditions, Cauchy data, rotation and/or Killing symmetries, etc. Physically important BH solutions are characterized by certain horizons and singularity, or regularity,  properties. The Schwarzschild BH metric consists of one of the most important examples of vacuum solutions in GR. It is defined by the condition $\Upsilon =0$ when $g(r)$ is a singular function on a radial coordinate $r$ (for a spherical system of space coordinates). It was constructed by using a diagonal ansatz, which transforms the Einstein equations into a system of nonlinear second-order ODEs. Such a diagonal metric has a maximum of 4 nontrivial coefficients determined by two integration constants.
%%%%%%%

\vskip5pt
One of the fundamental properties of GR is that certain singularity theorems \cite{penrose65,hawking73} can be proved for  very general and quite reasonable assumptions about a matter source $\Upsilon .$ To exclude the presence of naked singularities, many researchers also accept also the so-called weak cosmic censorship conjecture \cite{penrose69}. This means that the vacuum BH solutions describe the end of the collapse of matter in GR. In explicit form, the physical properties of such models were analyzed and tested theoretically and observationally using certain classes of diagonalizable metrics (by coordinate or frame transforms) constructed as solutions of the corresponding systems of nonlinear ODEs.
%%%%%%%

\vskip5pt  To construct off-diagonal solutions (in exact forms or with decompositions on a small parameter) of nonlinear systems (\ref{einststand}) consists of a very sophisticated geometric, analytic and numerical problem.\footnote{A pseudo-Riemannian metric is generic off-diagonal if it can't be diagonalized by coordinate transforms on a finite spacetime region.} The main technical issue is that using the geometric variables $(g,\nabla )$  the physically important systems of nonlinear PDEs can't be  decoupled and integrated (i.e. can't be solved) in certain general off-diagonal forms and with dependence of coefficients on all spacetime coordinates. For many decades in the past, the approach with diagonalizable and higher symmetry ansatz for metrics (which reduces nonlinear PDEs to some integrable ODEs) had very important physical motivations: Almost all observational and experimental data in cosmology and astrophysics pointed to a higher order homogeneity and isotropy of the Universe and spherical (for some cases, rotoid) symmetries of BH candidates etc.
%%%%%%%

\vskip5pt The "Orthodox" approach to GR with priorities of diagonal solutions was challenged after the discovery of the late-time cosmic acceleration \cite{riess98,perlmutter99}. A new "Protestant" era resulted in extensive research on many MGTs \cite{sotiriou10,nojiri11,capo11,clifton12,harko14}; see also \cite{copeland06}, on dark energy (DE) and dark matter (DM) physics; and on geometric and quantum information flows \cite{partner06}. During the last 30 years, we elaborated on the anholonomic frame and connection deformation method (AFCDM) as a general geometric and analytic method for constructing generic
off-diagonal solutions in GR and MGTs, see recent reviews \cite{vacaruplb16,vbubuianu17,partner02}.
%%%%%%%

\vskip5pt In GR, the main reason to study off-diagonal solutions is that general classes of exact or parametric solutions are described by 6 independent coefficients (from 10 ones for a symmetric metric tensor on a Lorentz manifold). For such solutions, we have 2additional degrees of freedom (compared to the maximum of 4 independent coefficients of a diagonal metric ansatz), which allows us to describe a new nonlinear physics and cosmology defined by off-diagonal gravitational and (effective) matter field interactions. We can follow the Orthodox paradigm of GR on selecting physical important solutions, at least partially: The main postulates are not modified but the  geometric and analytic methods of constructing exact and parametric solutions are generalized for solving directly physically important systems of  nonlinear PDEs.  New classes of  generic off-diagonal solutions are used for
elaborating models of nonlinear classical and quantum theories, locally anisotropic thermodynamics, and modelling in an effective form the DE and DM physics. A series of our works \cite{partner06,vacaruplb16,vbubuianu17,partner02,sv11} was devoted to constructing and analyzing physical properties of various classes of generic off-diagonal solutions (in GR and various MGTs). Such solutions describe nontrivial gravitational vacuum configurations, for instance, of solitonic types; pattern-creating and/or quasi-periodic structures; models with locally anisotropic polarizations of physical constants; or generating moving BHs, black ellipsoid/torus configurations with deformed horizons; and nonholonomic wormholes, WHs.
%%%%%%%

\vskip5pt The contributions of off-diagonal terms of metrics can be modelled in equivalent geometric forms by using respective distortions of linear connections, 
\begin{equation}
\nabla \rightarrow D=\nabla +Z.  \label{distr1}
\end{equation}%
In GR, a $D$ can be considered as an auxiliary metric-affine connection determined by a distortion tensor $Z.$ All geometric objects are defined on the same Lorentz manifold determined by $(g,\nabla )$ as solutions of  (\ref{einststand}). This results in effective sources $\ ^{e}\Upsilon $ encoding respective distortions of $Ric$ and $\ \Upsilon .$ This way, the Einstein equations (\ref{einststand}) can be expressed in an equivalent form,%
\begin{equation}
Ric[g,D]=\Upsilon +\ ^{e}\Upsilon ,  \label{einstdist}
\end{equation}%
using new geometric objects ($Ric[g,D],$ $\Upsilon \lbrack g,D]$ and $\ ^{e}\Upsilon \lbrack g,D]$) defined by the same metric structure $g$ as for 
$\nabla (g).$\footnote{Details on explicit constructions will be provided in the next section, see formulas (\ref{twocon}) and further constructions.} We can define such a canonical (in brief, hat connection) $\widehat{D}=\widehat{D}(g,\widehat{Z})$ adapted to a nonholonomic 2+2 spacetime decomposition, when the system of nonlinear PDEs (\ref{einstdist}) decouple in certain general forms for a class of off-diagonal metrics $g.$ This is the main result of the AFCDM, 
which allows us to decouple and construct in certain general forms generic off-diagonal solutions for certain general parameterizations of $\Upsilon +\
^{e}\Upsilon $ stated as (effective) generating sources. Respective details and proofs are presented in \cite{partner06,vacaruplb16,vbubuianu17,partner02,sv11}. After certain classes of off-diagonal solutions were found in general forms for $(g,\widehat{D})$, we can impose additional constraints, $\widehat{Z}\rightarrow 0$, to extract LC configurations and generate off-diagonal solutions in terms of $(g,\nabla ).$ Here we note that general decoupling properties with off-diagonal metrics can't be proven if we work from the very beginning with $\nabla ,$ or arbitrary $D.$ The main "geometric trick" is to use $\widehat{D}$  to construct general classes of solutions. Then, we can search if necessary for additional nonholonomic constraints or distortions which allow to generate solutions for $(g,\nabla ),$ or other types $(g,D).$ Not all such off-diagonal solutions may have (known) physical importance. A rigorous analysis is necessary for any special case of physical models constructed for an explicit class of off-diagonal spacetimes.
%%%%%%%

\vskip5pt Off-diagonal terms and effective sources $\ ^{e}\Upsilon $ can circumvent the singularity theorems in similar forms as known in GR for certain types of matter $\Upsilon $. Such details on diagonal configurations are presented in \cite{casadio25} and references therein. Both types of sources may allow the collapse to form an event horizon without leading to a singularity. For diagonal metrics, we can generate regular, i.e. completely non-singular BHs, which result in a second (Cauchy) horizon. This is particularly problematic \cite{poisson90} (for recent studies, see \cite{casadio23,casadio25}), which motivates the strong cosmic censorship conjecture.
%%%%%%%

\vskip5pt  In this paper, we do not study the fundamental problem on how singularities in GR are formed and the associated problem with the Cauchy horizons. The focus is on the construction and analysis of off-diagonal deformations of regular BHs. This is possible by applying the AFCDM \cite{partner06,vacaruplb16,vbubuianu17,partner02,sv11} and employing still reasonable forms of (effective) sources $(\ ^{e}\Upsilon )$ $\Upsilon .$ The
off-diagonal interactions are generic nonlinear and reflect a specific nonholonomic gravitational dynamics. If non-singular configurations  form during a collapse scenario, the effects associated with the Cauchy horizon of diagonal metrics seem to be unstable. But what about off-diagonal and nonholonomic scenarios? Should they also consist  for instance, a transitory state in the eventual formation of a singularity? One of the goals of this work is to investigate at least parametrically (with graviational polarization functions and for certain ellipsoid-type) off-diagonal deformations of regular BHs, how richer vacuum and non-vacuum structures could modify the formation of singularities.
%%%%%%%

\vskip5pt  The questions mentioned above require a detailed study of off-diagonal contributions in the inner BH regions, with possible deformations of horizons and effective polarization of physical constants. Most analyzed cases in GR, exhibit a simple internal geometry, which is consistent with the supposed higher symmetry of collapse scenarios \cite{hawking73,misner73,wald82}. We shall prove that off-diagonal interactions may modify the extremely simplicity of interior configurations modelled by diagonal solutions. This is compatible with the results of \cite{ovalle24} that the interior may be not extreme symple because the event horizons may form before the singularity appears. Such possibilities are real in the conditions stated by the weak cosmic censorship conjecture \cite{penrose69}.
In our approach, we investigate the alternative sources $\ ^{e}\Upsilon $ and $\Upsilon $ for exterior off-diagonal region of the Schwarzschild BH in GR.
%%%%%%%

\vskip5pt Because the generic off-diagonal solutions in GR do not involve, in general, certain horizon, duality, or holographic configurations, the
Bekenstein-Hawking thermodynamic paradigm \cite{bek2,haw2} is not applicable.  In a series of our works \cite{svnonh08,gheorghiuap16,partner06} (see references therein), we elaborated on relativistic and nonholonomic generalizations of thermodynamic variables defined by G. Perelman \cite{perelman1}. Such an alternative geometric flow approach can be used for characterizing the fundamental properties of off-diagonal solutions in GR and MGTs. We emphasize that in our works, we do not attempt to formulate and prove relativistic or non-Riemannian generalizations of the Poincar\`{e}-Thurston conjecture 
\cite{perelman1,hamilton82}. In rigorous mathematical forms, the proofs of (already)  Perelman's theorem are presented in monographs 
\cite{monogrrf1,monogrrf2,monogrrf3}. In those classical works and monographs, relativistic or metric-affine generalizations, related to GR and MGTs, of
the Ricci flow theory were not considered. For non-Riemannian geometries, this is a very difficult and undetermined task, for instance, in the case of
nonassociative and noncommutative geometric flows \cite{partner06}. This is because an infinite number of nonassociative/ noncommutative differential
and integral calculi can be defined which do not allow to unify the approaches and formulate, for instance, nonassociative versions of the
Poincar\`{e}-Thurston conjecture. Nevertheless, the sophisticated systems of nonlinear PDEs describing the relativistic geometric flows, with distortions
modeling geometric flow evolution of the (modified) Einstein equations, can be decoupled and solved in certain off-diagonal forms by applying the AFCDM.
The concept of W-entropy introduced for Ricci flows in \cite{perelman1} can be generalized and modified in various forms. This allows us to characterize
thermodynamically very general classes of off-diagonal solutions in GR and MGTs.
%%%%%%%

\vskip5pt  The aim of this work is four-fold as stated for respective sections: First, we outline the AFCDM for generating off-diagonal deformations of regular BH solutions (the objective, Obj1, for section \ref{sec2}). Second, we analyze the issue of off-diagonal interactions and the inside of BHs (Obj2, section \ref{sec3}). Section \ref{sec4} \ (Obj3) is devoted to a study of regular BHs and their off-diagonal deformations. Then, in section \ref{sec5} (Obj4), we study some applications of the relativistic geometric flow theory for characterizing off-diagonal deformations of regular BH solutions. We show how to define and compute the G. Perelman thermodynamic variables for respective classes of nonholonomic deformed regular Schwarzschild BH configurations. We conclude the results and perspectives in section \ref{sec6}.
%%%%%%%

\vskip5pt  
Finally, we emphasize that in this work we follow an abstract geometric approach to mathematical relativity. Technical results with frame and coordinate indices and geometric tedious proofs for generating solutions are omitted. Respective details are provided in \cite{partner06,vacaruplb16,vbubuianu17,partner02,sv11} (on decoupling and integrating in general form physically important systems of nonlinear PDEs) and, respectively, \cite{svnonh08,gheorghiuap16,partner06}, on relativistic and nonholonomic geometric flows and generalized G. Perelman's thermodynamics. Appendix \ref{appendixa} contains necessary formulas on other types of generating functions and gravitational polarizations including a technique for computing small parametric off-diagonal deformations. A brief introduction to the theory of relativistic geometric flow equations, and related formulas for nonholonomic Ricci solitons modelling Einstein spaces, is provided in Appendix \ref{appendixb}.
%%%%%%%

\section{Nonholonomic 2+2 splitting and connection distortions in GR}

\label{sec2}We outline necessary concepts and definitions from the geometry
of Lorentz manifolds enabled with conventional nonholonomic 2+2 splitting
and distortions of the linear connection structures. A general off-diagonal
ansatz with one time-like Killing symmetry is introduced. It allows us to
generate quasi-stationary solutions in GR and respective nonholonomic
deformations. Corresponding nonlinear symmetries and LC conditions are
explained and motivated. This geometric formalism for decopuling and
integrating (modified) Einstein equations can be extended for higher
dimensions and nonassociative and noncommutative, supersymmetric, nonmetric
and various types of other MGTs, see recent reviews of results and
references therein \cite{partner06,partner02}. In this section, the abstract
geometric and certain index and coefficient formulas are stated for 4-d
Lorentz spacetime manifolds.

\subsection{Off-diagonal metrics and auxiliary connections}

In this subsection, we consider such a nonholonomic dyadic formulation of
GR, which allows a general decoupling and integration of the Einstein
equations (\ref{einstdist}). Our system of notations consists of a modification
of conventions from \cite{hawking73,misner73,wald82} considered in general
form in \cite{partner06,partner02}. For convenience, we shall dub certain
necessary definitions and formulas in footnotes.

\subsubsection{Metric structures adapted to nonholonomic dyadic decompositions}

The GR theory is formulated in standard geometric form for Lorentz manifolds
(i.e. spacetimes) determined by geometric data $(V,\mathbf{g})$. In such an
approach, $V$ is a 4-d pseudo-Riemannian manifold of necessary smooth
(differentiability) class enabled with a symmetric metric tensor of
signature $(+++-),$ 
\begin{equation}
\mathbf{g}=g_{\alpha ^{\prime }\beta ^{\prime }}(u)e^{\alpha ^{\prime
}}\otimes e^{\beta ^{\prime }}.  \label{mst}
\end{equation}%
The tensor product $\otimes $ in (\ref{mst}) is used for some general
co-frames $e^{\alpha ^{\prime }}$ which are dual to frame bases $e_{\alpha
^{\prime }}.$\footnote{%
The Einstein convention on summarizing "up-low" repeating indices is assumed
if a contrary condition is not stated. We shall use different types of
indices (primed, underlined, etc.) to distinguish certain coordinates or frame
decompositions. On any coordinate neighborhood $U\subset V,$ we can
introduce a conventional $2+2$ splitting via local coordinates labeled as $%
u=\{u^{\alpha }=(x^{i},y^{a})\}.$ This way, we consider h-coordinates, $%
x=(x^{i}),$ and v-coordinates, $y=(y^{a}),$ for respective indices $%
j,k,...=1,2$ and $a,b,c,...=3,4,$ when $\alpha ,\beta ,...=1,2,3,4.$ To
compute the coefficients of geometric objects on $V,$ we can use local
coordinate bases and, respectively, co-bases, $e_{\alpha }=\partial _{\alpha
}=\partial /\partial u^{\beta }$ and $e^{\beta }=du^{\beta }.$ In general,
we can use arbitrary frames (tetrads) defined as $e_{\alpha ^{\prime }}=e_{\
\alpha ^{\prime }}^{\alpha }(u)e_{\alpha }$ and $e^{\alpha ^{\prime
}}=e_{\alpha \ }^{\ \alpha ^{\prime }}(u)e^{\alpha }$ ( such (co) bases are
orthonormal if $e_{\alpha \ }^{\ \alpha ^{\prime }}e_{\ \alpha ^{\prime
}}^{\beta }=\delta _{\alpha }^{\beta },$ where $\delta _{\alpha }^{\beta }$
is called the Kronecker symbol).}

For our purposes, we introduce a class of so-called N--elongated
(equivalently, N-adapted) local bases as certain partial derivative
operators, $\mathbf{e}_{\nu },$ and dual (co-) bases and instead of standard
differentials, $\mathbf{e}^{\mu }.$ \ We can consider on $V$ a set of
coefficients $\mathbf{N}(u)=\{N_{i}^{a}(x,y)\}$ and define such linear on $%
N_{i}^{a}$ local frames: 
\begin{eqnarray}
\mathbf{e}_{\nu } &=&(\mathbf{e}_{i},e_{a})=(\mathbf{e}_{i}=\partial
/\partial x^{i}-\ N_{i}^{a}(u)\partial /\partial y^{a},\ e_{a}=\partial
_{a}=\partial /\partial y^{a}),\mbox{ and  }  \label{nader} \\
\mathbf{e}^{\mu } &=&(e^{i},\mathbf{e}^{a})=(e^{i}=dx^{i},\ \mathbf{e}%
^{a}=dy^{a}+\ N_{i}^{a}(u)dx^{i}).  \label{nadif}
\end{eqnarray}%
Such bases are nonholonomic (equivalently, anholonomic, i.e.
non-integrable). For instance, a basis (\ref{nader}) satisfies nonholonomic
relations 
\begin{equation}
\lbrack \mathbf{e}_{\alpha },\mathbf{e}_{\beta }]=\mathbf{e}_{\alpha }%
\mathbf{e}_{\beta }-\mathbf{e}_{\beta }\mathbf{e}_{\alpha }=W_{\alpha \beta
}^{\gamma }\mathbf{e}_{\gamma },  \label{nonholr}
\end{equation}%
when the (antisymmetric) anholonomy coefficients are computed 
\begin{equation}
W_{ia}^{b}=\partial _{a}N_{i}^{b},W_{ji}^{a}=\mathbf{e}_{j}\left(
N_{i}^{a}\right) -\mathbf{e}_{i}(N_{j}^{a}).  \label{anhcoef}
\end{equation}
A N-adapted base $\mathbf{e}_{\alpha }\simeq \partial _{\alpha }=\partial
/\partial u^{\alpha }$ is holonomic if and only if all anholonomic
coefficients (\ref{anhcoef}) are zero. For holonomic configurations, usual
partial derivatives $\partial _{\alpha }$ can be obtained for certain
coordinate transforms. In curved spacetime coordinates, for holonomic bases,
the coefficients $N_{j}^{a}$ may be non-zero even if all $W_{\alpha \beta
}^{\gamma }=0.$

The above 2+2 spacetime decompositions can be defined in coordinate-free form
using nonholonomic 2+2 distributions on $V$. For this, in global form, we
can consider a Whitney sum $\ \mathbf{N}:\ TV=hV\oplus vV$ for the tangent
Lorentz bundle $TV,$ which defines a \textit{nonlinear connection} structure
(\textit{N-connection}).  E. Cartan \cite{cartan35} used N-connections in
coordinate forms (on tangent bundles) and the concept was defined in
rigorous mathematical form in \cite{ehresmann55}.\footnote{%
We emphasize that in this work, the N-connections are considered on 4-d
Lorentz manifolds. This is quite different from a similar geometric object
in Finsler geometry when the N-connections are defined by splitting of type $%
TTV=hTV\oplus vTV.$ More details and references can be found in \cite%
{partner06,partner02}.} A N-connection $\mathbf{N}$ can be defined by a
nonholonomic distribution stated by a set of coefficients $\mathbf{N}%
(u)=\{N_{i}^{a}(x,y)\}$ used in formulas (\ref{nader}) and (\ref{nadif})
and when in local coordinate form 
\begin{equation}
\mathbf{N}=N_{i}^{a}(x,y)dx^{i}\otimes \partial /\partial y^{a}.
\label{nconcoef}
\end{equation}%
Hereafter, we shall omit priming, underlying, overlying, etc. of indices if
that does not result in ambiguities. \ We shall use "boldface" symbols to
emphasize that certain spaces or geometric objects are enabled (or adapted)
with (to) an N-connection structure.

The concept of \textit{nonholonomic Lorentz manifold} $\mathbf{V}$ is used
and if such a spacetime is enabled with N-connection structure, which allows
to define N-adapted frames. Using general covariant frame and coordinate
transforms on $\mathbf{V}$, we can also consider general frame structures, $%
e_{\alpha }=(e_{i},e_{a})$ and $e^{\beta }=(e^{i},e^{a}),$ or certain
coordinate frames, $e_{\alpha }=\partial _{\alpha }$ and $e^{\beta
}=du^{\beta }$ etc. The terms of distinguished, d, geometric objects
(d-objects, d-tensors, d-connections etc.) are used in our works emphasize
that the geometric constructions, formulas, equations and respective
solutions are performed/ defined on a nonholonomic Lorentz manifold. For
d-objects, the N-adapted coefficients are computed with respect to some
frames (\ref{nader}) and (\ref{nadif}), or their tensor products etc. Ffor
instance, we can write a d--vector as $\mathbf{X}=(hX,vX)$.

On $\mathbf{V}$, a spacetime metric $\mathbf{g}$ (\ref{mst}) can be written
equivalently as a d--metric or, respectively, as an off-diagonal metric (in
local coordinate form), 
\begin{eqnarray}
\ \mathbf{g} &=&(hg,vg)=\ g_{ij}(x,y)\ e^{i}\otimes e^{j}+\ g_{ab}(x,y)\ 
\mathbf{e}^{a}\otimes \mathbf{e}^{b}  \label{dm} \\
&=&\underline{g}_{\alpha \beta }(u)du^{\alpha }\otimes du^{\beta }.
\label{cm}
\end{eqnarray}%
In these formulas, $hg=\{\ g_{ij}\}$ and $\ vg=\{g_{ab}\},$ when
off-diagonal coefficients in (\ref{cm}) are computed if we introduce the
coefficients of (\ref{nadif}) into (\ref{dm}) with a corresponding
regrouping for a coordinate dual basis. This way, we obtain a formula
relating the N-connection coefficients to corresponding off-diagonal and
diagonal coefficeints of metrics written in local coordinate bases, 
\begin{equation}
\underline{g}_{\alpha \beta }=\left[ 
\begin{array}{cc}
g_{ij}+N_{i}^{a}N_{j}^{b}g_{ab} & N_{j}^{e}g_{ae} \\ 
N_{i}^{e}g_{be} & g_{ab}%
\end{array}%
\right] .  \label{ansatz}
\end{equation}%
We note that a metric $\mathbf{g}=\{\underline{g}_{\alpha \beta }\}$ (\ref%
{ansatz}) (equivalently (\ref{dm})) is generic off--diagonal if there are
nonzero anholonomy coefficients $W_{\alpha \beta }^{\gamma }$ (\ref{anhcoef}%
). For 4-d (pseudo) Riemannian spaces, such a matrix can not be diagonalized
on a finite spacetime region using only coordinate transforms. The above
formulas for d-metrics and off-diagonal metrics can be defined for any
prescribed set of coefficients $N_{i}^{a},$ i.e. for any nonholonomic
distributijns. Parameterizations of type (\ref{ansatz}) are used, for
instance, in Kaluza-Klein gravity when $N_{j}^{e}=A_{j}^{e}$ are identified
as certain gauge fields after compactification on $y$-coordinates (usually,
there are considered higher dimension spacetimes). In GR, the N-coefficients
can be arbitrary (gravitational) ones defined by certain off-diagonal
solutions of the Einstein equations.

\subsubsection{Distinguished metric-affine structures and canonical dyadic
(d-) variables}

For nonholonomic Lorentz manifold $\mathbf{V}$ defined by crtain d-metric
and N-connection structures ($\mathbf{g}$, $\mathbf{N}$), we can define
certain classes of linear connections which are, or not, N-adapted. Here we
note that a LC connection $\nabla $ does not preserve under parallel
transports an h- and v-splitting and is not adapted. To perform a covariant
differential and integral calculus in N-adapted form (in GR) we need a more
special class of distinguished connections, d-connections.

A \textbf{\ d--connection} $\mathbf{D}=(hD,vD)$ can be introduced in GR as a
linear connection preserving under parallelism the N--connection splitting.
It defines a N--adapted covariant derivative $\mathbf{D}_{\mathbf{X}}\mathbf{%
Y}$ of a d--vector field $\mathbf{Y}=hY+vY$ in the direction of a d--vector $%
\mathbf{X}=hX+vC.$\ With respect to (\ref{nader}) and (\ref{nadif}), any $%
\mathbf{D}_{\mathbf{X}}\mathbf{Y}$ can be computed as in \cite{misner73} but
when the N-adapted coefficients involve respective h- and v-indices, 
\begin{equation}
\mathbf{D}=\{\mathbf{\Gamma }_{\ \alpha \beta }^{\gamma }=(L_{jk}^{i},\acute{%
L}_{bk}^{a};\acute{C}_{jc}^{i},C_{bc}^{a})\},\mbox{ where }hD=(L_{jk}^{i},%
\acute{L}_{bk}^{a})\mbox{ and }vD=(\acute{C}_{jc}^{i},C_{bc}^{a}),
\label{hvdcon}
\end{equation}%
see details in \cite{vacaruplb16,vbubuianu17,partner02,partner06}.

Any d--connection $\mathbf{D}$ is characterized by three fundamental
geometric d-objects: 
\begin{eqnarray}
\mathcal{T}(\mathbf{X,Y}):= &&\mathbf{D}_{\mathbf{X}}\mathbf{Y}-\mathbf{D}_{%
\mathbf{Y}}\mathbf{X}-[\mathbf{X,Y}],\mbox{ torsion d-tensor,  d-torsion};
\label{fundgeom} \\
\mathcal{R}(\mathbf{X,Y}):= &&\mathbf{D}_{\mathbf{X}}\mathbf{D}_{\mathbf{Y}}-%
\mathbf{D}_{\mathbf{Y}}\mathbf{D}_{\mathbf{X}}-\mathbf{D}_{\mathbf{[X,Y]}},%
\mbox{ curvature d-tensor, d-curvature};  \notag \\
\mathcal{Q}(\mathbf{X}):= &&\mathbf{D}_{\mathbf{X}}\mathbf{g,}%
\mbox{nonmetricity d-fiels, d-nonmetricity}.  \notag
\end{eqnarray}%
The coefficients of such geometric d-objects can be computed in N-adapted
form by introducing $\mathbf{X}=\mathbf{e}_{\alpha }$ and $\mathbf{Y}=%
\mathbf{e}_{\beta },$ defined by (\ref{nader}), and considering a
h-v-splitting $\mathbf{D}=\{\mathbf{\Gamma }_{\ \alpha \beta }^{\gamma }\}$ (%
\ref{hvdcon}) into above formulas. We omit cumbersome formulas for such
coefficients but note that they can be parameterized in such forms: 
\begin{eqnarray}
\mathcal{T} &=&\{\mathbf{T}_{\ \alpha \beta }^{\gamma }=\left( T_{\
jk}^{i},T_{\ ja}^{i},T_{\ ji}^{a},T_{\ bi}^{a},T_{\ bc}^{a}\right) \};
\label{fundgeomc} \\
\mathcal{R} &\mathbf{=}&\mathbf{\{R}_{\ \beta \gamma \delta }^{\alpha }%
\mathbf{=}\left( R_{\ hjk}^{i}\mathbf{,}R_{\ bjk}^{a}\mathbf{,}R_{\ hja}^{i}%
\mathbf{,}R_{\ bja}^{c}\mathbf{,}R_{\ hba}^{i},R_{\ bea}^{c}\right) \mathbf{%
\};}  \notag \\
\ \mathcal{Q} &=&\mathbf{\{Q}_{\ \alpha \beta }^{\gamma }=\mathbf{D}^{\gamma
}\mathbf{g}_{\alpha \beta }=(Q_{\ ij}^{k},Q_{\ ij}^{c},Q_{\ ab}^{k},Q_{\
ab}^{c})\}.  \notag
\end{eqnarray}

Any geometric data $\left( \mathbf{V},\mathbf{N},\mathbf{g,D}\right) $
define a nonholonomic, or N-adapted, metric-affine structure (equivalently,
metric-affine d-structure). In general, such a d-metric and a d-connection
are stated independently. Using $\mathbf{g,}$ we can define and compute in
standard form $\nabla (\mathbf{g})$ and (similarly to (\ref{distr1})) define
a N-adapted distortion relation $\mathbf{D=\nabla +Z.}$ A d-tensor $\mathbf{Z%
}$ is defied differently in various types of MGTs.\footnote{%
If we work with a LC connection $\nabla ,$ defined by the conditions $%
\mathcal{Q}[\nabla ]=\nabla \mathbf{g=0}$ and $\mathcal{T}[\nabla ]=0,$ we
obtain standard abstract and index formulas for GR. For tensors and not
d-tensors, we use not boldface symbols and consider functional dependencies $%
[\nabla ]$ or abstract left labels, for instance, for the curvature tensor
we have $\mathcal{R}[\nabla ]=\ _{\nabla }\mathcal{R}=\{R_{\ \beta \gamma
\delta }^{\alpha }=\mathbf{\ }_{\nabla }R_{\ \beta \gamma \delta }^{\alpha
}\}.$ Explicit formulas for the coefficients (\ref{fundgeomc}) of different
types of d-connections and linear connections are provided in \cite%
{vbubuianu17,partner02}.}

\subsubsection{The Einstein equations in canonical d-variables}

Using a d-metric structure $\mathbf{g}$ (\ref{dm}) adapted to a N-connection 
$\mathbf{N}$ (\ref{nconcoef}), we can define in GR different types of linear
connections. For our purposes, we shall consider two important linear
connection structures: 
\begin{equation}
(\mathbf{g,N})\rightarrow \left\{ 
\begin{array}{cc}
\mathbf{\nabla :} & \mathbf{\nabla g}=0;\ _{\nabla }\mathcal{T}=0,\ 
\mbox{\
the LC--connection }; \\ 
\widehat{\mathbf{D}}: & \widehat{\mathbf{Q}}=0;\ h\widehat{\mathcal{T}}=0,v%
\widehat{\mathcal{T}}=0,\ hv\widehat{\mathcal{T}}\neq 0,%
\mbox{ the canonical
d-connection}.%
\end{array}%
\right.  \label{twocon}
\end{equation}%
The LC connection is the standard one in (pseudo) Riemannian geometry but it
does not allow a general decoupling of (modified) gravitational equations
for generic off-diagonal metrics. In our works \cite%
{partner06,vacaruplb16,vbubuianu17,partner02,sv11}, we found that it
possible to elaborate the AFCDM and construct very general classes of
solutions in GR and MGTs using an (auxiliary) "hat" connection, i.e. a
d-connection, $\widehat{\mathbf{D}}.$ In this case, the ditortion relation (%
\ref{distr1}) transforms into a canonical distortion relation, 
\begin{equation}
\widehat{\mathbf{D}}[\mathbf{g}]=\nabla \lbrack \mathbf{g}]+\widehat{%
\mathcal{Z}}[\mathbf{g}],  \label{canondistrel}
\end{equation}%
where $\widehat{\mathcal{Z}}=\{\widehat{\mathbf{Z}}_{\ \alpha \beta
}^{\gamma }\}$ is the canonical distortion d-tensor. We argue that the GR
theory can be defined equivalently using both types of geometric data $[%
\mathbf{g},\nabla ]$ and (or) $[\mathbf{g},\mathbf{N},\widehat{\mathbf{D}}].$
The priority of hat variables allow to decouple and integrate the Einstein
equations, the the nontrivial N-connection structure $\mathbf{N}$ absorbing
in a sense the off-diagonal terms in $\mathbf{g}=\{\underline{g}_{\alpha
\beta }\}$ (\ref{ansatz}). We note that the distortions (\ref{canondistrel})
involve a nontrivial canonical d-torsion structure, $\widehat{\mathcal{T}}=\{%
\widehat{\mathbf{T}}_{\ \alpha \beta }^{\gamma }\}$ (stated in (\ref{twocon}%
)). Such a torsion is different from that used, for instance, in the
Einstein-Cartan theory \cite{cartan35,misner73} because $\widehat{\mathcal{T}%
}$ and $\widehat{\mathcal{Z}}$ are nonholonomically induced by certain
N-connection and d-metric coefficients. This distorts the fundamental
geometric objects and physical equations in GR but do not need additional
(spin-like) sources for the canonical d-torsion. We can include the
distortions of the Ricci d-tensor determined by (\ref{canondistrel}) as
certain effective matter sources in the Einstein equations for $[\mathbf{g}%
,\nabla ]$. An alternative variant for extracting LC configurations is to
impose additional constraints on the generating and integration functions
(see next subsection) for respective off-diagonal solutions which result in
zero distortion d-tensors, 
\begin{equation}
\widehat{\mathbf{Z}}=0,\mbox{ which is equivalent to }\ \widehat{\mathbf{D}}%
_{\mid \widehat{\mathcal{T}}=0}=\nabla .  \label{lccond}
\end{equation}

The N-adapted coefficients (\ref{hvdcon}) of a canonical d-connection $%
\widehat{\mathbf{D}}=\{\widehat{\mathbf{\Gamma }}_{\ \alpha \beta }^{\gamma
}=(\widehat{L}_{jk}^{i},\widehat{L}_{bk}^{a},\widehat{C}_{jc}^{i},\widehat{C}%
_{bc}^{a})\}$ (\ref{twocon}) can be computed for a d--metric $\mathbf{g}%
=[g_{ij},g_{ab}]$ (\ref{dm}) using N--elongated partial derivatives (\ref%
{nader}).\footnote{%
We present respective coefficient formulas: 
\begin{eqnarray}
\widehat{L}_{jk}^{i} &=&\frac{1}{2}g^{ir}(\mathbf{e}_{k}g_{jr}+\mathbf{e}%
_{j}g_{kr}-\mathbf{e}_{r}g_{jk}),\widehat{L}_{bk}^{a}=e_{b}(N_{k}^{a})+\frac{%
1}{2}g^{ac}(\mathbf{e}_{k}g_{bc}-g_{dc}\ e_{b}N_{k}^{d}-g_{db}\
e_{c}N_{k}^{d}),  \notag \\
\widehat{C}_{jc}^{i} &=&\frac{1}{2}g^{ik}e_{c}g_{jk},\ \widehat{C}_{bc}^{a}=%
\frac{1}{2}g^{ad}(e_{c}g_{bd}+e_{b}g_{cd}-e_{d}g_{bc}).  \notag
\end{eqnarray}%
In a similar form, we can compute the coefficients of an LC connection $%
\nabla =\{\Gamma _{\ \alpha \beta }^{\gamma }\},$ see general coefficient or
N-adapted formulas in \cite{misner73,partner02}. The N-adapted coefficients
of the canonical distortion d-tensor in (\ref{canondistrel}) can be found as 
$\widehat{\mathbf{Z}}=\{\widehat{\mathbf{Z}}_{\ \alpha \beta }^{\gamma }=%
\widehat{\mathbf{\Gamma }}_{\ \alpha \beta }^{\gamma }-\Gamma _{\ \alpha
\beta }^{\gamma }\}.$Using $\widehat{\mathbf{\Gamma }}_{\ \alpha \beta
}^{\gamma }$\ instead of the coefficients of a general d-connection $\mathbf{%
D=\{\Gamma }_{\ \alpha \beta }^{\gamma }\}$, we can compute the N-adapted
coefficients (\ref{fundgeom}) of the canonical fundamental d--objects.} For
such canonical dyadic variables, we use hat symbols and write $\widehat{%
\mathcal{R}}=\{\widehat{\mathbf{R}}_{\ \beta \gamma \delta }^{\alpha }=(%
\widehat{R}_{\ hjk}^{i},\widehat{R}_{\ bjk}^{a},...)\},\ \widehat{\mathcal{T}%
}=\{\widehat{\mathbf{T}}_{\ \alpha \beta }^{\gamma }=(\widehat{T}_{\ jk}^{i},%
\widehat{T}_{\ ja}^{i},...)\},$ for $\widehat{\mathcal{Q}}=\{\widehat{%
\mathbf{Q}}_{\gamma \alpha \beta }=(\widehat{Q}_{kij}=0,\widehat{Q}%
_{kab}=0)=0.$ The canonical distortion relation for linear connections (\ref%
{canondistrel}) allow us to compute respective canonical distortions of the
fundamental geometric d-objects for $\nabla $. Such formulas relate, for
instance, two different curvature tensors, $\ _{\nabla }\mathcal{R}=\{\
_{\nabla }R_{\ \beta \gamma \delta }^{\alpha }\}$ and $\ \widehat{\mathcal{R}%
}=\{\widehat{\mathbf{R}}_{\ \beta \gamma \delta }^{\alpha }\}$ etc.

The canonical Ricci d-tensor $\widehat{\mathbf{R}}ic$ is defined in standard
form, for instance, by contractiing the the 1st and 4th indices of the
canonical curvature d-tensor,%
\begin{equation}
\widehat{\mathbf{R}}ic=\{\widehat{\mathbf{R}}_{\ \beta \gamma }:=\widehat{%
\mathbf{R}}_{\ \beta \gamma \alpha }^{\alpha }\}.  \label{criccidt}
\end{equation}%
Because of nonholonomic structure, this d-tensor is not symmetric, $\widehat{%
\mathbf{R}}_{\ \beta \gamma }\neq \widehat{\mathbf{R}}_{\ \gamma \beta }.$
The canonical scalar curvature is also defined in standard form,%
\begin{equation}
\widehat{R}sc:=\mathbf{g}^{\alpha \beta }\widehat{\mathbf{R}}_{\ \alpha
\beta }.  \label{criccidsc}
\end{equation}%
Respectively, the (nonholonomic) canonical Einstein d-tensor is defined by
using (\ref{criccidt}) and (\ref{criccidsc}), 
\begin{equation}
\widehat{\mathbf{E}}n:=\widehat{\mathbf{R}}ic-\frac{1}{2}\mathbf{g}\widehat{R%
}sc=\{\widehat{\mathbf{R}}_{\ \beta \gamma }-\frac{1}{2}\mathbf{g}_{\ \beta
\gamma }\widehat{R}sc\}.  \label{criccdsc}
\end{equation}%
We can also introduce in N-adapted form necessary types of energy-momentum
sources $\widehat{\mathbf{T}}_{\alpha \beta }$ as in GR, when $\nabla
\rightarrow \widehat{\mathbf{D}},$ and postulate in abstract geometric form
(as in \cite{misner73}) the Einstein equations in hat variables, 
\begin{equation}
\widehat{\mathbf{E}}n_{\alpha \beta }=\varkappa \widehat{\mathbf{T}}_{\alpha
\beta }.  \label{einstceq1}
\end{equation}%
In these formulas, the constant $\varkappa $ can be related to the Newton
gravitational constant. Such nonholonomic equations can be equivalent to the
standard Einstein ones in GR if $\widehat{\mathbf{T}}_{\alpha \beta }$ is
constructed in a way to include as sources the energy-momentum tensors for
matter (in N-adapted bases) but also the distortion terms coming from $%
En_{\alpha \beta }[\nabla ]$ and $T_{\alpha \beta }[\nabla ].$ Similar
arguments hold for $\ ^{e}\widehat{\Upsilon }:=$ $\ ^{e}\Upsilon \lbrack 
\mathbf{g,}\widehat{\mathbf{D}}]$ as we explained above for (\ref{einstdist}%
). Respective distortions of fundamental geometric d-objects have to be
computed using (\ref{canondistrel}). Alternatively, we obtain an equivalence
of (\ref{einstceq1}) with the gravitational field equations in GR (for
instance, in the standard form (\ref{einststand})) if we \ impose
additionally the nonholonomic constraints (\ref{lccond}) for extracting LC
configurations.

We emphasize that the gravitational equations (\ref{einstceq1}) can be
proven also in N-adapted variational form. Let us consider a gravitational
Lagrange density, $\ ^{g}L(\widehat{\mathbf{R}}sc)$ (as in GR with $\
^{g}L(R)$) and a matter fields Lagrange density, $\ ^{m}L(\varphi ^{A},%
\mathbf{g}_{\beta \gamma }).$ The stress-energy d-tensor of matter fields $%
\varphi ^{A}$ (labelled by a general index $A$) is defined and computed as
in GR but using nonholonomic dyadic decompositions, 
\begin{equation}
\mathbf{T}_{\alpha \beta }=-\frac{2}{\sqrt{|\mathbf{g}_{\mu \nu }|}}\frac{%
\delta (\ ^{m}L\sqrt{|\mathbf{g}_{\mu \nu }|})}{\delta \mathbf{g}^{\alpha
\beta }}.  \label{emdt}
\end{equation}%
Using the trace $T:=\mathbf{g}^{\alpha \beta }\mathbf{T}_{\alpha \beta },$
we construct hat sources of Ricci d-tensors, $\widehat{\mathbf{Y}}[\mathbf{g,%
}\widehat{\mathbf{D}}]\simeq \{\mathbf{T}_{\alpha \beta }-\frac{1}{2}\mathbf{%
g}_{\alpha \beta }T\}$ as in (\ref{einststand}). In physical theories, one
considers more general $\ ^{m}L,$ for instance, depending on some
covariant/spinor derivatives etc. For simplicity, in this work, we sider
functionals of type $\ ^{m}L(\varphi ^{A},\mathbf{g}_{\beta \gamma })$ not
depending on covariant derivatives.

We are able to decouple and integrate in general form (modified) Einstein
equations if we consider (effective) sources $\widehat{\mathbf{Y}}[\mathbf{g,%
}\widehat{\mathbf{D}}]=\{\Upsilon _{~\delta }^{\beta }(x,y)\}$ parameterized
in such forms: 
\begin{equation}
\widehat{\Upsilon }_{~\delta }^{\beta }=diag[\Upsilon _{\alpha }:\Upsilon
_{~1}^{1}=\Upsilon _{~2}^{2}=~^{h}\Upsilon (x^{k});\Upsilon
_{~3}^{3}=\Upsilon _{~4}^{4}=~^{v}\Upsilon (x^{k},y^{a})].  \label{esourc}
\end{equation}%
Such coefficients are defined with respect to N-adapted frames (\ref{nader})
and (\ref{nadif}). This types of parameterizations can be obtained for some
general classes of energy-momentum tensors by using respective
frame/coordinate transforms (if such conditions are not satisfied for a $%
\Upsilon _{\beta \delta }).$ We also consider that $\widehat{\Upsilon }%
_{~\delta }^{\beta }$ can be related by certain distortion relations to
source $\Upsilon +\ ^{e}\Upsilon $ in (\ref{einstdist}). Such assumptions
allows us to construct generic off-diagonal solutions for certain classes of
nonholonomic transforms and constraints when the effective sources are
determined by \textbf{two generating sources} $\ ^{h}\Upsilon (x^{k})$ and $%
\ ^{v}\Upsilon (x^{k},y^{a})$. This defines a conventional phase space
configuration of (effective) matter sourses defined as nonholonomic
constraints (\ref{esourc}) on $\mathbf{T}_{\alpha \beta }$, possible
cosmological constant $\Lambda $ and for a more special splitting of
constants into h- and v-components. The nonholonomic constraints may involve
distortion d-tensors $\widehat{\mathbf{Z}}[\mathbf{g}]$ and other values
included in $\widehat{\mathbf{Y}}.$ Typically, such constrained equations
can be solved by considering decompositions on a small parameter (a physical
constant, or a deformation constant). ertheless, the conditions (\ref{esourc}%
) allow us to decouple and integrate in general explicit forms the geometric
flows and gravitational and matter field equations in many cases \cite%
{vacaruplb16,vbubuianu17,partner02}.

The Einstein equations (\ref{einstceq1}) can be written in a form similar to
(\ref{einstdist}), which is more convenient for decoupling and integrating:
, 
\begin{eqnarray}
\widehat{\mathbf{R}}_{\ \ \beta }^{\alpha } &=&\widehat{\mathbf{\Upsilon }}%
_{\ \ \beta }^{\alpha },  \label{cdeq1} \\
\widehat{\mathbf{T}}_{\ \alpha \beta }^{\gamma } &=&0,%
\mbox{ if we extract
LC configuations with }\nabla .  \label{lccond1}
\end{eqnarray}%
In this system of nonlinear PDEs, we use generating sources $\widehat{%
\mathbf{\Upsilon }}_{\ \ \beta }^{\alpha }=[\ ^{h}\Upsilon \delta _{\ \
j}^{i},\ ^{v}\Upsilon \delta _{\ \ b}^{a}]$ (\ref{esourc}), the the
equations (\ref{lccond1}) are equivalent to (\ref{lccond}). We have also an
induced nonholonomic d-torsion $\widehat{\mathcal{T}}=\{\widehat{\mathbf{T}}%
_{\ \alpha \beta }^{\gamma }[\mathbf{g,N,}\widehat{\mathbf{D}}]\}$ which is
defined as in (\ref{fundgeom}) and must be stated to zero for LC
configurations.

The conservation laws for (\ref{cdeq1}) are typical ones for nonholonomic
systems.In general, $\widehat{\mathbf{D}}^{\beta }\widehat{\mathbf{E}}_{\ \
\beta }^{\alpha }\neq 0$ and $\widehat{\mathbf{D}}^{\beta }\widehat{\mathbf{%
\Upsilon }}_{\ \ \beta }^{\alpha }\neq 0,$ which is different from the
properties of the Einstein and energy-momentum tensors written in standard
form in GR. Such non-zero covariant divergences are typical for nonholonomic
systems. This is similar to the nonholonomic mechanics when the conservation
laws are not standard ones. In nonholonomic canonical variables, using
distortions of connections (\ref{canondistrel}), we can rewrite (\ref{cdeq1}%
) in terms of $\nabla ,$ when $\nabla ^{\beta }E_{\ \ \beta }^{\alpha
}=\nabla ^{\beta }T_{\ \ \beta }^{\alpha }=0.$

\subsection{Quasi-stationary ansatz for generating off-diagonal solutions}

Let us consider a consider an ansatz for a d-metric of type (\ref{dm}), 
\begin{eqnarray}
\mathbf{\hat{g}} &=&g_{i}(x^{k})dx^{i}\otimes dx^{i}+h_{3}(x^{k},y^{3})%
\mathbf{e}^{3}\otimes \mathbf{e}^{3}+h_{4}(x^{k},y^{3})\mathbf{e}^{4}\otimes 
\mathbf{e}^{4},  \notag \\
&&\mathbf{e}^{3}=dy^{3}+w_{i}(x^{k},y^{3})dx^{i},\ \mathbf{e}%
^{4}=dy^{4}+n_{i}(x^{k},y^{3})dx^{i},  \label{dmq}
\end{eqnarray}%
when the hat variables are used for the the d-metric coefficients, $\widehat{%
\mathbf{g}}_{\alpha \beta }=[\widehat{g}_{ij}(x^{\kappa }),\widehat{g}%
_{ab}(x^{\kappa },y^{3})],$ and the N-connection coefficients, $\widehat{N}%
_{i}^{3}=w_{i}(x^{k},y^{3})$ and $\widehat{N}_{i}^{4}=n_{i}(x^{k},y^{3})$
(such functions can be of necessary smooth class). The same ansatz in
off-diagonal form (\ref{ansatz}) can be written as 
\begin{eqnarray}
\widehat{g} &=&\widehat{g}_{\alpha \beta }(u)du^{\alpha }\otimes du^{\beta },%
\mbox{ when }  \label{qeltorsoffd} \\
\widehat{g}_{\alpha \beta } &=&\left[ 
\begin{array}{cccc}
e^{\psi }+(w_{1})^{2}h_{3}+(n_{1})^{2}h_{4} & w_{1}w_{2}h_{3}+n_{1}n_{2}h_{4}
& w_{1}h_{3} & n_{1}h_{4} \\ 
w_{1}w_{2}h_{3}+n_{1}n_{2}h_{4} & e^{\psi }+(w_{2})^{2}h_{3}+(n_{2})^{2}h_{4}
& w_{2}h_{3} & n_{2}h_{4} \\ 
w_{1}h_{3} & w_{2}h_{3} & h_{3} & 0 \\ 
n_{1}h_{4} & n_{2}h_{4} & 0 & h_{4}%
\end{array}%
\right] .  \notag
\end{eqnarray}
Parametrizations of type (\ref{dmq}) and (\ref{qeltorsoffd}) can be obtained
using some frame or coordinate transforms even, in general, such a $\mathbf{%
\hat{g}}(u)$ may depend on all spacetime coordinates. We use a hat label for 
$\mathbf{\hat{g}}$ to emphasize that (\ref{dmq}) is \textit{quasi-stationary}
(with a Killing symmetry \ on $\partial _{4}=\partial _{t}$), i.e. the
coefficients do not depend on the time-like coordinate $y^{4}=t.$ Similar
ansatz can be considered for d-metrics with other orders of space coordinates%
$,$ for instance, $(x^{1},y^{3},x^{2})$, when the v-coordinate is $x^{2}$
instead of $y^{3}.$ Corresponding space coordinates can be spherical,
cylindric, toroid and other types.

Tedious computations in N-adapted frames show that the Einstein equations in
canonical d-variable (\ref{cdeq1}) are solved by a quass-stationary ansatz (%
\ref{dmq}) if respective coefficents are computed as follows: 
\begin{eqnarray}
g_{1}(x^{k}) &=&g_{2}(x^{k})=e^{\psi (x^{k})}%
\mbox{ are determined by
solutions of 2-d Poisson equations: }\psi ^{\bullet \bullet }+\psi ^{\prime
\prime }=\ ^{h}\Upsilon ,  \label{dmqs} \\
h_{3}(x^{k},y^{3}) &=&-\frac{1}{4h_{4}}\left( \frac{\Psi ^{\ast }}{\
^{v}\Upsilon }\right) ^{2}=-\left( \frac{\Psi ^{\ast }}{2\ \ ^{v}\Upsilon }%
\right) ^{2}\left( h_{4}^{[0]}-\int dy^{3}\frac{[\Psi ^{2}]^{\ast }}{4\ \
^{v}\Upsilon }\right) ^{-1},  \notag \\
\ h_{4}(x^{k},y^{3}) &=&h_{4}^{[0]}-\int dy^{3}[\Psi ^{2}]^{\ast }/4(\
^{v}\Upsilon );\mbox{ and }  \notag
\end{eqnarray}

\begin{eqnarray}
w_{i}(x^{k},y^{3}) &=&\partial _{i}\ \Psi /(\Psi )^{\ast },  \label{nconqs}
\\
n_{k}(x^{k},y^{3}) &=&\ _{1}n_{k}+\ _{2}n_{k}\int dy^{3}\ \frac{h_{3}}{|\
h_{4}|^{3/2}}=\ _{1}n_{k}+\ _{2}n_{k}\int dy^{3}\left( \frac{\Psi ^{\ast }}{%
2\ ^{v}\Upsilon }\right) ^{2}|\ h_{4}|^{-5/2},  \notag \\
&=&\ _{1}n_{k}+\ _{2}n_{k}\int dy^{3}\left( \frac{\Psi ^{\ast }}{2\
^{v}\Upsilon }\right) ^{2}\left\vert h_{4}^{[0]}-\int dy^{3}[\Psi
^{2}]^{\ast }/4\ ^{v}\Upsilon \right\vert ^{-5/2}.  \notag
\end{eqnarray}%
In these formulas, $\Psi (x^{k},y^{3})$ is a generating function; $\
^{h}\Upsilon (x^{k})$ and $\ ^{v}\Upsilon (x^{k},y^{3})$ are generating
sources (\ref{esourc}); and $h_{4}^{[0]}=h_{4}^{[0]}(x^{k}),$ $\ _{1}n_{k}=\
_{1}n_{k}(x^{i})$ and $\ _{2}n_{k}=\ _{2}n_{k}(x^{i})$ are integration
functions. This is a typical terminology from the theory of nonlinear PDEs
(see \cite{kramer03} for applications in mathematical relativity). In our
works, we use brief notations of partial derivatives, for instance, $%
\partial _{1}q(u^{\alpha })=q^{\bullet },$ $\partial _{2}q(u^{\alpha
})=q^{\prime }$, $\partial _{3}q(u^{\alpha })=q^{\ast }$ and $\partial
_{4}q(u^{\alpha })=q^{\diamond }$ (for quasi-stationary solutions, we do not
need $\diamond $).

References \cite{sv11,vacaruplb16,vbubuianu17,partner02} (with reviews and
applications of the AFCDM) contain technical details and proofs that the
coefficents (\ref{dmqs}) and (\ref{nconqs}) can be used for generating
off-diagonal solutions in GR and MGTs. Such solutions are generic
off-diagonal because, in general, the corresponding anholonomy coefficients (%
\ref{anhcoef}) are not zero. We need to impose additional nonholonomic
constraints to make zero the nonholonomic torsion, $\widehat{\mathbf{T}}_{\
\alpha \beta }^{\gamma }=0$ (\ref{lccond1}), and extract LC configurations
(how to solve this issue if necessary, see next subsection). In the
mentiones works, the AFCDM is also developed for generating locally
anisotropic cosmological solutions (with certain duality properties to the
quasi-stationary ones) and for more general ansatz than (\ref{dmq}) and (\ref%
{qeltorsoffd}) with more general dependencies on all spacetime coordinates.
We omit such details in this work.

The corresponding quadratic element for quasi-stationary off-diagonal
solutions in canonical dyadic variables can be written in the form%
\begin{eqnarray}
d\widehat{s}^{2} &=&e^{\psi (x^{k},\ ^{h}\Upsilon
)}[(dx^{1})^{2}+(dx^{2})^{2}]+\frac{[\Psi ^{\ast }]^{2}}{4(\ ^{v}\Upsilon
)^{2}\{g_{4}^{[0]}-\int dy^{3}[\Psi ^{2}]^{\ast }/4(\ ^{v}\Upsilon )\}}%
(dy^{3}+\frac{\partial _{i}\Psi }{\Psi ^{\ast }}dx^{i})^{2}+  \label{qeltors}
\\
&&\{g_{4}^{[0]}-\int dy^{3}\frac{[\Psi ^{2}]^{\ast }}{4(\ ^{v}\Upsilon )}%
\}\{dt+[\ _{1}n_{k}+\ _{2}n_{k}\int dy^{3}\frac{[(\Psi )^{2}]^{\ast }}{4(\
^{v}\Upsilon )^{2}|g_{4}^{[0]}-\int dy^{3}[\Psi ^{2}]^{\ast }/4(\
^{v}\Upsilon )|^{5/2}}]dx^{k}\}.  \notag
\end{eqnarray}%
We note that such a canonical form of d-metrics is written for the
conditions that $\Psi ^{\ast }\neq 0$ and $\ ^{v}\Upsilon \neq 0.$ If such
conditions are not satisfied, we need more special considerations, see \cite%
{sv11,vbubuianu17,partner02} and references therein. Any d-metric (\ref%
{qeltors}) decouples and solves the (modified) Einstein equations (\ref%
{cdeq1}) in a general form (for a correspodning system of nonlinear PDEs)
not cutting other classes of exact and parametric solutions as in the case
of diagonal ansatz resulting in ODEs and integration constancts. The generic
off-diagonal solutions are determined by 6 degrees of freedom and determined
by generating functions and generating sources which have to be prescribed
following certain physical arguments. The integration functions have to be
defined from certain boundary/ asymptotic conditions and other symmetry
conditions.

\subsection{Nonlinear symmetries and effective cosmological constants}

The quasi-stationary solutions (\ref{qeltors}) are described by important
nonlinear symmetries which allow us to transform the generating functions
and effective sources into other types of generating functions with
effective cosmological constants. By tedious computations \cite%
{vbubuianu17,sv11}, we can prove that such solutions admit a changing of the
generating data, $(\Psi ,\ \ ^{v}\Upsilon )\leftrightarrow (\Phi ,\
^{v}\Lambda =const\neq 0)$, when an effective v-cosmological constant$\
^{v}\Lambda $ may be different from an effectife h-cosmological constant $\
^{h}\Lambda .$ In GR, we can consider $\ ^{h}\Lambda =\ ^{v}\Lambda =$ $%
\Lambda ,$ and consider $\Lambda \neq 0.$ More special cases when generic
off-diagonal gravitational interactions result in an effective $\Lambda =0$
need a more special analysis.

We assume that for certain nonlinear transforms, the quasi-stationary
solutions $\mathbf{\hat{g}}[\Psi ]$ (\ref{qeltors}) of (\ref{cdeq1}) can be
expressed in an equivalent class of solutions of 
\begin{equation}
\widehat{\mathbf{R}}_{\ \ \beta }^{\alpha }[\Phi ]=\Lambda \mathbf{\delta }%
_{\ \ \beta }^{\alpha }  \label{cdeq1a}
\end{equation}%
with another generating function $\Phi (x^{k},y^{3})$ . Such a system of
nonlinear PDEs involves an effective cosmological constant $\Lambda .$ The
generating data $(\Phi ,\ \Lambda ),$ or $(\Psi ,\ \ ^{v}\Upsilon )$ can be
chosen, for instance, to describe stationary DE, related to an effective $%
\Lambda ,$ and DM effects (determined by effective matter) in in
accelerating cosmology. The conditions that the quasi-stationary
configurations (\ref{qeltors}) are described equivalently by
quasi-stationary solutions of (\ref{cdeq1a}) are stated by such differential
or integral equations (this can be checked by straightforward computations): 
\begin{eqnarray}
\frac{\lbrack \Psi ^{2}]^{\ast }}{\ ^{v}\Upsilon } &=&\frac{[\Phi
^{2}]^{\ast }}{\Lambda },\mbox{ which can be integrated as  }
\label{ntransf1} \\
\Phi ^{2} &=&\ \Lambda \int dy^{3}(\ ^{v}\Upsilon )^{-1}[\Psi ^{2}]^{\ast }%
\mbox{ and/or }\Psi ^{2}=\Lambda ^{-1}\int dy^{3}(\ ^{v}\Upsilon )[\Phi
^{2}]^{\ast }.  \label{ntransf2}
\end{eqnarray}%
Using (\ref{ntransf1}), we can simplify the formulas (\ref{dmqs}) by writing 
$h_{4}=h_{4}^{[0]}-\frac{\ \Phi ^{2}}{4\ \Lambda }.$ This allows us to
express all d-metric and N-connection coefficients for (\ref{qeltors}) in
terms of new generating data, when $(\Psi )^{\ast }/\ ^{v}\Upsilon $ can be
re-defined for $(\Phi ,\Lambda ).$The formulas in (\ref{ntransf1}) and (\ref%
{ntransf2}) are written in a new form:%
\begin{equation*}
\frac{\Psi (\ \Psi )^{\ast }}{\ ^{v}\Upsilon }=\frac{(\Phi ^{2})^{\ast }}{%
2\Lambda }\mbox{ and }\ \Psi =|\Lambda |^{-1/2}\sqrt{|\int dy^{3}\
^{v}\Upsilon \ (\Phi ^{2})^{\ast }|}.
\end{equation*}%
Introducing $\Psi $ from the above second equation in the first equation, we
re-define $\Psi ^{\ast }$ in terms of generating data $(\ ^{v}\Upsilon ,\Phi
,\Lambda ),$ when 
\begin{equation*}
\frac{\Psi ^{\ast }}{\ ^{v}\Upsilon }=\frac{[\Phi ^{2}]^{\ast }}{2\sqrt{|\
\Lambda \int dy^{3}(\ ^{v}\Upsilon )[\Phi ^{2}]^{\ast }|}}.
\end{equation*}

Using above formulas for nonlinear symmetries, the quadratic element for
quasi-stationary solutions (\ref{qeltors}) can be written in an equivalent
form 
\begin{eqnarray}
d\widehat{s}^{2} &=&\widehat{g}_{\alpha \beta }(x^{k},y^{3},\Phi ,\Lambda
)du^{\alpha }du^{\beta }=e^{\psi (x^{k})}[(dx^{1})^{2}+(dx^{2})^{2}]
\label{offdiagcosmcsh} \\
&&-\frac{\Phi ^{2}[\Phi ^{\ast }]^{2}}{|\Lambda \int dy^{3}\ ^{v}\Upsilon
\lbrack \Phi ^{2}]^{\ast }|[h_{4}^{[0]}-\Phi ^{2}/4\Lambda ]}\{dy^{3}+\frac{%
\partial _{i}\ \int dy^{3}\ ^{v}\Upsilon \ [\Phi ^{2}]^{\ast }}{\
^{v}\Upsilon \ [(\ \Phi )^{2}]^{\ast }}dx^{i}\}^{2}-  \notag \\
&&\{h_{4}^{[0]}-\frac{\Phi ^{2}}{4\Lambda }\}\{dt+[\ _{1}n_{k}+\
_{2}n_{k}\int dy^{3}\frac{\Phi ^{2}[\Phi ^{\ast }]^{2}}{|\Lambda \int
dy^{3}\ ^{v}\Upsilon \lbrack \Phi ^{2}]^{\ast }|[h_{4}^{[0]}-\Phi
^{2}/4\Lambda ]^{5/2}}]\}.  \notag
\end{eqnarray}%
In this formula, the generating functions are parameterized $\psi (x^{k})$
and $\Phi (x^{k}y^{3});$ the generating sources are $^{h}\Upsilon (x^{k})$
and $\ ^{v}\Upsilon (x^{k},y^{3});$ and the effective cosmological constant
is $\ \Lambda .$ Integration functions$\ _{1}n_{k}(x^{j}),\ _{2}n_{k}(x^{j})$
and $g_{4}^{[0]}(x^{k}),$ are also considered as we explained for the
previous off-diagonal ansatz. We can say that the quasi-stationary solutions
represented in the form (\ref{offdiagcosmcsh}) "disperse" into respective
off-diagonal forms the prescribed generating data $(\Psi ,\ ^{v}\Upsilon ).$
They are re-defined into another ones, $(\Phi ,\ \Lambda )$, with effective
cosmological constant, but the contributions of a generating source $\
^{v}\Upsilon $ are not completely transformed into a cosmological constant $%
\Lambda .$ The coefficients of d-metrics $\mathbf{\hat{g}}[\Phi ,\
^{v}\Upsilon ,\Lambda ]$ (\ref{offdiagcosmcsh}) keep certain memory about
effective and real matter fields $\ $stated in $\mathbf{\hat{g}}[\Psi ,\ \
^{v}\Upsilon ]$ (\ref{qeltors}). In Appendix \ref{appendixa}, we present
formulas for other types of generating functions and gravitational
polarizations which are important for study physical properties of
off-diagonal solutions.

The nonlinear symmetries and the possibility of introducing an effective $%
\Lambda $ simplifies the method of computing G. Perelman thermodynamic
variables as we show at the end of section \ref{sec5}. \ We also conclude
that any quasi-stationary solution expressed in a form (\ref{qeltors}) or (%
\ref{offdiagcosmcsh}) possess important nonlinear symmetries of type (\ref%
{ntransf1}) and (\ref{ntransf2}) which are trivial or do not exist for
diagonal ansatz. They describe a new off-diagonal nonlinear gravitational
physics which can't be detected if we restrict our research in GR only to
diagonal ansatz.

\subsection{ Solutions for the LC conditions}

To generate exact and parametric solutions in GR we have to solve additional
anholonomic constraints of type (\ref{lccond}) or (\ref{lccond1}) and
extract LC configurations with $\nabla $. This is possible for restricting
the classes of generating and integrations functions in respective
off--diagonal quasi-stationary solutions of (\ref{cdeq1}) or (\ref{cdeq1a})
constructed for the canonical d--connection $\widehat{\mathbf{D}}.$

Straightforward computations for quasi-stationary configurations show that
all d-torsion coefficients vanish if the coefficients of the N--adapted
frames and the $v$--components of d--metrics are solutions of such
equations: 
\begin{eqnarray}
\ w_{i}^{\ast }(x^{i},y^{3}) &=&\mathbf{e}_{i}\ln \sqrt{|\
h_{3}(x^{i},y^{3})|},\mathbf{e}_{i}\ln \sqrt{|\ h_{4}(x^{i},y^{3})|}%
=0,\partial _{i}w_{j}=\partial _{j}w_{i}\mbox{ and }n_{i}^{\ast }=0;  \notag
\\
n_{k}(x^{i}) &=&0\mbox{ and }\partial _{i}n_{j}(x^{k})=\partial
_{j}n_{i}(x^{k}).  \label{zerot1}
\end{eqnarray}%
The solutions for the off-diagonal cooefficents and respective $w$- and $n$%
-functions depend on the class of vacuum or non--vacuum metrics. If we
prescribe a generating function $\Psi =\check{\Psi}(x^{i},y^{3})$ which
satisfy the condtions $[\partial _{i}\check{\Psi}]^{\ast }=\partial _{i}%
\check{\Psi}^{\ast },$ we can solve the equations for $w_{j}$ from (\ref%
{zerot1}). In explicit form, we can take $\ ^{v}\Upsilon =const,$ or to
express the effective source as a functional $\ ^{v}\Upsilon (x^{i},y^{3})=\
\ ^{v}\Upsilon \lbrack \check{\Psi}].$ At the next step, we can solve the
third conditions $\partial _{i}w_{j}=\partial _{j}w_{i}$ if we chose a
generating function $\ \check{A}=\check{A}(x^{k},y^{3})$ and define 
\begin{equation*}
w_{i}(x^{i},y^{3})=\check{w}_{i}(x^{i},y^{3})=\partial _{i}\ \check{\Psi}/(%
\check{\Psi})^{\ast }=\partial _{i}\check{A}(x^{i},y^{3}).
\end{equation*}%
The equations for $n$-functions in (\ref{zerot1}) are solved if $%
n_{i}(x^{k})=\partial _{i}[\ ^{2}n(x^{k})].$

For the subclasses of above generating data, we can write the quadratic
element for quasi-stationary solutions with zero canonical d-torsion in a
form similar to (\ref{qeltors}), 
\begin{eqnarray}
d\check{s}^{2} &=&\check{g}_{\alpha \beta }(x^{k},y^{3})du^{\alpha
}du^{\beta }  \label{qellc} \\
&=&e^{\psi }[(dx^{1})^{2}+(dx^{2})^{2}]+\frac{[\check{\Psi}^{\ast }]^{2}}{%
4(\ ^{v}\Upsilon \lbrack \check{\Psi}])^{2}\{h_{4}^{[0]}-\int dy^{3}[\check{%
\Psi}]^{\ast }/4\ ^{v}\Upsilon \lbrack \check{\Psi}]\}}\{dy^{3}+[\partial
_{i}(\check{A})]dx^{i}\}^{2}  \notag \\
&&+\{h_{4}^{[0]}-\int dy^{3}\frac{[\check{\Psi}^{2}]^{\ast }}{4(\
^{v}\Upsilon \lbrack \check{\Psi}])}\}\{dt+\partial _{i}[\
^{2}n]dx^{i}\}^{2}.  \notag
\end{eqnarray}%
Similar constraints on generation functions as in (\ref{zerot1}) can be
solved in non-explicit forms by imposing certain nonholonomic constraints on
the N-connection and d-metric coefficients and effective sources. They can
be solved also in explicit form for some small parameters. Correspondingly
re-defined nonlinear symmetries allow us to extract LC configurations for
all classes of quasi-stationary solutions analyzed in this section. The
generating data for generic off-diagonal metrics with trivial canonical
d-torsion are lableled in general form as $(\check{\Psi}(x^{i},y^{3}),\
^{v}\Upsilon \lbrack \ \check{\Psi}],\check{A}(x^{k},y^{3})).$

\section{Off-diagonal interactions inside the BHs}

\label{sec3} We study how generic off-diagonal interactions modify the
inside of BHs using gravitational polarization functions as defined in
Appendix \ref{assgravpol}.

\subsection{Small parametric gravitational polarizations of BH constants}

In this subsection, we analyze the possibility of locally anisotropic
polarizations of constants (and related off-diagonal deformations) inside a
BH.

\subsubsection{Prime metrics inside BHs}
As a prime metric, we consider a static and spherical symmetric ansatz 
\begin{equation}
d\mathring{s}^{2}=+f^{-1}(r)dr^{2}+r^{2}d\Omega ^{2}-f(r)dt^{2}  \label{pm1}
\end{equation}%
for a metric written in Kerr-Schield form \cite{casadio24,casadio25,kramer03}, see also references therein on previous results for BH solutions. In our approach, the spherical coordinates are labeled: $x^{1}=r,x^{2}=\theta ,y^{3}=\varphi $ and $y^{4}=t,$ where $d\Omega =d\Omega (\theta ,\varphi )$
is determined by angular coordinates. In (\ref{pm1}), the metric function is
defined as 
\begin{equation}
f(r)=\left\{ 
\begin{array}{ccc}
f^{-}:=1-\frac{2m(r)}{r}, & \mbox{ for } & 0<r\leq \mathring{h}; \\ 
f^{+}:=1-\frac{2\mathring{m}}{r}, & \mbox{ for } & r>\mathring{h},%
\end{array}%
\right.  \label{metricf}
\end{equation}%
where the horizon $r=\mathring{h}=2\mathring{m}$ is defined by the
coordinate singularity, where $\mathring{m}$ is the ADM mass \cite{misner73}
associated with a point-like singularity at the center $r=0$. The mass
function in (\ref{metricf}) is defined respectively 
\begin{equation}
m(r)=\left\{ 
\begin{array}{ccc}
\mathring{m} & \mbox{ for }0<r, & 
\mbox{ we obtain the Schwarzschild
solution}; \\ 
m(\mathring{h})=\mathring{h}/2=\mathring{m} & \mbox{ for }r>\mathring{h}. & 
\end{array}%
\right.  \label{massf}
\end{equation}%
In these formulas, we consider that as for any function $m(\mathring{h}%
):=m(r)_{|r=\mathring{h}}$ and the units with $c=1$ and $\varkappa =8\pi
G_{N},$ where $G_{N}$ is the Newton's gravitational constant.

A prime diagonal metric (\ref{pm1}) with metric and mass functions,
respectively, (\ref{metricf}) and (\ref{massf}) define an exact solution of
the Einstein equations (\ref{einststand}) with the energy-momentum tensor (%
\ref{emdt}) is defined 
\begin{equation}
\ _{m}T_{\ \beta }^{\alpha }(r)=diag[p_{r}(r)=-\frac{2m^{\bullet }}{%
\varkappa r^{2}},p_{\theta }(r)=-\frac{m^{\bullet \bullet }}{\varkappa r^{2}}%
,p_{\theta }(r),-\epsilon (r)=-\frac{2m^{\bullet }}{\varkappa r^{2}}],
\label{emtensint}
\end{equation}%
where $m^{\bullet }=\partial _{1}m=\partial _{r}m.$ This ~$\ _{m}T_{\ \beta
}^{\alpha }$ (\ref{emtensint}) can be derived in variational form under the
condition that the matter Lagrange density $\ ^{m}L=0$ for $r>\mathring{h}$
(and this condition holds inside the horizon when $0<r<\mathring{h}).$ We
use a left up or low label (on convenience) to emphasize that such physical
objects are determined by a mass function (\ref{massf}).

The Bianchi identities on Lorentz manifolds, $\nabla _{\alpha }E^{\alpha \beta
}=0,$impose the continuous equation $\epsilon ^{\bullet }=\frac{2}{r}%
(p_{r}(r)-p_{\theta }(r)).$ So, the energy density $\epsilon $ decreases
monotonically from the center (with $\epsilon ^{\bullet }<0$) if $%
p_{r}(r)<p_{\theta }(r).$ The continuity condition of values (\ref{metricf})
and (\ref{massf}) accros the horizon $r=\mathring{h}$ with $m(\mathring{h})=%
\mathring{m}$ is satisfied if%
\begin{equation}
m^{\bullet }(\mathring{h})=0\mbox{ and }\epsilon (\mathring{h})=p_{r}(%
\mathring{h})=0.  \label{continucond}
\end{equation}
Here we note that $p_{\theta }(r)$ can be discontinuous across $r=\mathring{h}.$

To apply the AFCDM and perform off-diagonal deformations without additional
coordinate singularities we can perform some frame transforms of \ (\ref{pm1}) to a parametrization with trivial N-connection coefficients $\mathring{N} _{i}^{a}=$ $\mathring{N}_{i}^{a}(u^{\alpha }(r,\theta ,\varphi ,t))$ and $\mathring{g}_{\beta }(u^{j}(r,\theta ,\varphi ),u^{3}(r,\theta ,\varphi )).$
We can introduce new coordinates 
\begin{equation}
u^{1}=x^{1}=r,u^{2}=\theta ,\mbox{ when }u^{3}=y^{3}(r,\theta ,\varphi
)=\varphi +\ ^{3}B(r,\theta ),u^{4}=y^{4}(r,\theta ,t)=t+\ ^{4}B(r,\theta ),
\label{newcoord}
\end{equation}
when 
\begin{eqnarray*}
\mathbf{\mathring{e}}^{3}(r,\theta ,y^{3}) &=&d\varphi =du^{3}+\mathring{N}%
_{i}^{3}(r,\theta )dx^{i}=du^{3}+\mathring{N}_{1}^{3}(r,\theta )dr+\mathring{%
N}_{2}^{3}(r,\theta )d\theta , \\
\mathbf{\mathring{e}}^{4}(r,\theta ,y^{3}) &=&dt=du^{4}+\mathring{N}%
_{i}^{4}(r,\theta )dx^{i}=du^{4}+\mathring{N}_{1}^{4}(r,\theta )dr+\mathring{%
N}_{2}^{4}(r,\theta )d\theta ,
\end{eqnarray*}%
are defined for $\mathring{N}_{i}^{3}(r,\theta ,(r,\theta ,y^{3}(r,\theta
,\varphi ))=-\partial \ ^{3}B/\partial x^{i}$ and $\mathring{N}%
_{i}^{4}(r,\theta ,(r,\theta ,y^{3}(r,\theta ,\varphi ))=-\partial \
^{4}B/\partial x^{i}.$ This way, we express  the quadratic line elements
for the above BH solutions (\ref{pm1}) as a prime d-metric, 
\begin{equation}
d\mathring{s}^{2}=\mathring{g}_{\alpha }(r,\theta ,y^{3}(r,\theta ,\varphi
))[\mathbf{\mathring{e}}^{\alpha }(r,\theta ,y^{3}(r,\theta ,\varphi ))]^{2},
\label{pm1a}
\end{equation}%
where $\mathring{g}_{1}=f^{-1}(r),\mathring{g}_{2}=r^{2},\mathring{g}%
_{3}=r^{2}\sin ^{2}\theta $ and $\check{g}_{4}=-f^{-1}(r).$ Such a prime
d-metric can be always diagonalized by coordinate tranforms if we chose
frame transforms with $W_{\alpha \beta }^{\gamma }[\mathring{N}_{i}^{a}]=0,$
see (\ref{anhcoef}) but $\mathring{N}_{i}^{a}\neq 0$ which is important
applying the AFCDM.

The above changing of coordinates for constructing the prime d-metric (\ref{pm1a}) introduces some "fictive" N-coefficients $\mathring{N}_{i}^{a}$ but for a holonomic basis $\mathbf{\mathring{e}}_{\alpha }=(\mathbf{\mathring{e}}_{i}=\partial _{i}-\mathring{N}_{i}^{a}(r,\theta )\partial _{a},\mathbf{%
\mathring{e}}_{a}=\partial _{a}),$ when $\partial _{i}=\partial /\partial
u^{i}$ and $\partial _{a}=\partial /\partial u^{a}$ as in (\ref{nader}).
Such formulas can be considered inside a horizon $r<\mathring{h}$ or outside
a horizon $r>\mathring{h},$ considering respective smooth/ singular
properties for ($\mathring{g}_{\alpha },\mathring{N}_{i}^{a}$)  when $r=%
\mathring{h}.$ We also assume that corresponding frame/coordinate transforms
express the energy-momentum tensor ~ (\ref{emtensint}) into a generating
source of type (\ref{esourc}), 
\begin{equation}
\ _{m}T_{\ \beta }^{\alpha }(r)\rightarrow \ _{m}\widehat{\mathbf{\Upsilon }}%
_{\ \ \beta }^{\alpha }=[\ _{m}^{h}\Upsilon (m(r),x^{k}(r,\theta ))\delta
_{\ \ j}^{i},\ _{m}^{v}\Upsilon (m(r),x^{k}(r,\theta ),y^{3}(r,\theta
,\varphi ))\delta _{\ \ b}^{a}].  \label{msourc}
\end{equation}%
The explicit functional formulas with $m(r)$ in (\ref{msourc}) depend on the
chosen system of reference/coordinates. Using the AFCDM, we can generate
quasi-stationary solutions of (\ref{cdeq1}) and effective sources 
\begin{equation}
\ \widehat{\mathbf{\Upsilon }}_{\ \ \beta }^{\alpha }=\ _{e}\widehat{\mathbf{%
\Upsilon }}_{\ \ \beta }^{\alpha }+\ _{m}\widehat{\mathbf{\Upsilon }}_{\ \
\beta }^{\alpha }=diag[\ ^{h}\Upsilon =\ _{e}^{h}\Upsilon +\
_{m}^{h}\Upsilon ,\ ^{v}\Upsilon =\ _{e}^{v}\Upsilon +\ _{m}^{v}\Upsilon ],
\label{esourcm}
\end{equation}%
where $\ _{e}\widehat{\mathbf{\Upsilon }}_{\ \ \beta }^{\alpha }=diag[\
_{e}^{h}\Upsilon ,\ _{e}^{v}\Upsilon ]$ encode quasi-stationary distortions
of the Ricci d-tensor for canonical distortions of linear connections (\ref%
{canondistrel}). If for constructing off-diagonal deformations of (\ref{pm1a}%
), we use generating functions $(\check{\Psi}(x^{i},y^{3}),\ ^{v}\Upsilon
\lbrack \ \check{\Psi}],\check{A}(x^{k},y^{3}))$ as in (\ref{qellc}), we
solve the LC conditions $\widehat{\mathbf{T}}_{\ \alpha \beta }^{\gamma }=0$
(\ref{lccond1}). Hereafter we shall work with canonical nonholonomic
generating data $(\Psi ,\ ^{v}\Upsilon =\ _{e}^{v}\Upsilon +\
_{m}^{v}\Upsilon ),$ or $(\Phi ,\ \Lambda )$ considering that LC
configurations can be always extracted as for d-metrics (\ref{qellc}) and
their equivalent re-definitions of generating data explained in Appendix \ref%
{assgravpol}.

\subsubsection{General parametric gravitational polarizations of BH constants}

Introducing generating e- and m-sources (\ref{esourcm}) into the
quasi-stationary ansatz (\ref{offdiagpolfr}), we construct off-diagonal new
classes of off-diagonal solutions of with $\eta $-polarization, 
\begin{eqnarray}
d\widehat{s}^{2} &=&\widehat{g}_{\alpha \beta }(x^{k},y^{3};\mathring{g}%
_{\alpha };\psi ,\eta _{4};\ \Lambda ,\ _{e}^{v}\Upsilon +\ _{m}^{v}\Upsilon
)du^{\alpha }du^{\beta }=e^{\psi (r,\theta )}[(dx^{1}(r,\theta
))^{2}+(dx^{2}(r,\theta ))^{2}]  \label{newbh} \\
&&-\frac{[\partial _{\varphi }(\eta _{4}\mathring{g}_{4})]^{2}}{|\int
d\varphi \ (\ _{e}^{v}\Upsilon +\ _{m}^{v}\Upsilon )\partial _{\varphi
}(\eta _{4}\ \mathring{g}_{4})|\ \eta _{4}\breve{g}_{4}}\{dy^{3}+\frac{%
\partial _{i}[\int d\varphi \ (\ _{e}^{v}\Upsilon +\ _{m}^{v}\Upsilon )\
\partial _{\varphi }(\eta _{4}\mathring{g}_{4})]}{(\ _{e}^{v}\Upsilon +\
_{m}^{v}\Upsilon )\partial _{\varphi }(\eta _{4}\mathring{g}_{4})}%
dx^{i}\}^{2}  \notag \\
&&+\eta _{4}\mathring{g}_{4}\{dt+[\ _{1}n_{k}(r,\theta )+\
_{2}n_{k}(r,\theta )\int d\varphi \frac{\lbrack \partial _{\varphi }(\eta
_{4}\mathring{g}_{4})]^{2}}{|\int d\varphi \ (\ _{e}^{v}\Upsilon +\
_{m}^{v}\Upsilon )\partial _{\varphi }(\eta _{4}\mathring{g}_{4})|\ (\eta
_{4}\mathring{g}_{4})^{5/2}}]dx^{k}\}^{2}.  \notag
\end{eqnarray}%
This family of quasi-stationary solutions is determined by a generating
function $\eta _{4}(x^{k},y^{3})=\eta _{4}(r,\theta ,\varphi )$ and
respective integration functions, $\ _{1}n_{k}(r,\theta )$ and $\
_{2}n_{k}(r,\theta ).$ The function $\psi (r,\theta )$ is defined as a
solution of 2-d Poisson equation 
\begin{equation}
\partial _{11}^{2}\psi +\partial _{22}^{2}\psi =2(\ _{e}^{h}\Upsilon
(r,\theta )+\ _{m}^{h}\Upsilon (r,\theta )).  \label{poisson}
\end{equation}%
In the above formulas, we use re-definitions of the coordinates $(x^{k},y^{3})\rightarrow (r,\theta ,\varphi )$ (\ref{newcoord}) which allow us to symplify the formulas and understand how the primary metric (\ref{pm1}) and corresponding primary d-metric (\ref{pm1a}) are off-diagonally derormed
into target d-metrics (\ref{newbh}).

Let us speculate how locally anisotropic polarizations of masses can be
encoded in (\ref{pm1a}). We prescribe a new metric function with anisotropy
on $\theta ,$ when 
\begin{equation}
f(r)=\left\{ 
\begin{array}{ccc}
f^{-}:=1-\frac{2m(r)}{r}, & \mbox{ for } & 0<r\leq \mathring{h}; \\ 
f^{+}:=1-\frac{2\mathring{m}}{r}, & \mbox{ for } & r>\mathring{h},%
\end{array}%
\right. \Longrightarrow \tilde{f}(r,\theta )=\left\{ 
\begin{array}{ccc}
f^{-}:=1-\frac{2\tilde{m}(r,\theta )}{r}, & \mbox{ for } & 0<r\leq \mathring{%
h}; \\ 
f^{+}:=1-\frac{2\mathring{m}}{r}, & \mbox{ for } & r>\mathring{h},%
\end{array}%
\right. ,  \label{metrfa}
\end{equation}%
for some $\tilde{m}(r,\theta )\simeq m(r)$ if we neglect angular anisotropy.
Considering two h-polarizations functions $\eta _{i}(r,\theta )$ and a
v-polarization function $\eta _{4}(r,\theta ,\varphi ),$ we write 
\begin{eqnarray*}
g_{1} &=&g_{2}=\eta _{1}(r,\theta )f^{-1}(r)=\eta _{2}(r,\theta
)r^{2}=e^{\psi (r,\theta )}=\tilde{f}^{-1}(r,\theta )\mbox{ and } \\
h_{4} &=&\eta _{4}\mathring{g}_{4}=-\eta _{4}(r,\theta ,\varphi )f(r)=\tilde{%
f}(r,\theta )\tilde{\eta}_{4}(\varphi ),h_{3}=-\frac{[\partial _{\varphi }(%
\tilde{\eta}_{4})]^{2}}{|\int d\varphi \ (\ _{e}^{v}\Upsilon +\
_{m}^{v}\Upsilon )\partial _{\varphi }(\tilde{\eta}_{4})|\tilde{\eta}_{4}}.
\end{eqnarray*}%
Introducing these coefficients in (\ref{newbh}), we generate a locally
anisotropic off-diagonal polarization of the interior of the Schwarzschild
solution determined by polarization of the mass function/ constant $%
m(r)\rightarrow $ $\tilde{m}(r,\theta )$, $\ $with $f(r)\Longrightarrow 
\tilde{f}(r,\theta ),$ via $\tilde{f}(r,\theta )\tilde{\eta}_{4}(\varphi ).$
Such gravitational and generating source polarizations of the BH interior
are general ones if the values $\tilde{f}(r,\theta )$ and $\tilde{\eta}_{4}(\varphi )$ 
do not involve any small parameters. We still keep a formal horizon because
we use the same $f^{+}:=1-\frac{2\mathring{m}}{r}$ in (\ref{metricf}) and (%
\ref{metrfa}) but the general generating data $(\ _{e}^{v}\Upsilon +\
_{m}^{v}\Upsilon ),\tilde{m}(r,\theta )$ and $\tilde{\eta}_{4}(\varphi )$
change radically the interior structure of (former) BH configurations.

The locally anisotropic vacuum effects in the interior of a BH, which are  determined by a
d-metric with anisotropic vertical coordinate (\ref{esourcm}), are very
complex, and it is difficult to state well-defined and general conditions
when they preserve BH configurations. We need additional assumptions on
smoothness of generating and integration functions, their behaviour on the
prescribed horizon of the primary configurations (\ref{pm1}) and (\ref{pm1a}%
) to preserve for the target d-metrics (\ref{newbh}) the BH properties by
embedding into a non-trivial gravitational vacuum, for instance, of
solitonic type. A corresponding stability analysis is necessary. Such an
research was performed in our former works \cite{vi17,vbubuianu17} for
certain black ellipsoid, BE, configurations in string theory for some
explicitly generating and integrating data. This is necessary if we try to
construct stable target configurations. Here we note that non-stable
solutions may also have certain physical importance, for instance,
describing some evolution, or structure formation, for a period of time or
under certain temperature regimes.
%%%%

We also emphasize that any quasi-stationary d-metric (\ref{newbh}) can be characterized by nonlinear symmetries of type (\ref{nonlinsymrex}),  \begin{eqnarray}
\partial _{\varphi }[\Psi ^{2}] &=&-\int d\varphi \ \ (\ _{e}^{v}\Upsilon +\
_{m}^{v}\Upsilon )\partial _{\varphi }h_{4}  \label{nlims2} \\
&\simeq &-\int d\varphi \ (\ _{e}^{v}\Upsilon +\ _{m}^{v}\Upsilon )\partial
_{\varphi }(\eta _{4}\ \mathring{g}_{4})\simeq -\int d\varphi \ \ (\
_{e}^{v}\Upsilon +\ _{m}^{v}\Upsilon )\partial _{\varphi }[\zeta
_{4}(1+\kappa \ \chi _{4})\ \mathring{g}_{4}],  \notag \\
\Psi &=&|\ \Lambda |^{-1/2}\sqrt{|\int d\varphi \ \ (\ _{e}^{v}\Upsilon +\
_{m}^{v}\Upsilon )\ (\Phi ^{2})^{\ast }|},\Phi ^{2}=-4\ \Lambda h_{4}\simeq
-4\Lambda \eta _{4}\mathring{g}_{4}\simeq -4\Lambda \zeta _{4}(1+\kappa \chi
_{4})\mathring{g}_{4}.  \notag
\end{eqnarray}%
This allows us to introduce an effective cosmological constant $\Lambda $
which may be different from that considered for some quantum fluctuations,
or as in the original Einstein equations (typically denoted $\lambda $) for $%
\nabla .$ By additional assumptions, we can identify $\Lambda $ with $%
\lambda ,$ but even in such cases we do not eliminate the essential
nonlinear character of GR and MGTs which are  effectively modelled on Lorentz
manifolds. In this work, we can consider $\Lambda $ as an effective constant
which allows us to "smear" the nonlinear off-diagonal interactions by
changing equivalently the generating functions. This does not allow us to
eliminate from the formulas the generating sources (\ref{esourcm}%
) but emphasize the existence of certain types of nonlinear symmetries (\ref%
{nlims2}) for the off-diagonal interior of BH configurations. Using $\Lambda
,$ we can simplify the computation of the respective G. Perelman's thermodynamic
variables, see section \ref{sec5}.
%%%%%

If the prime metric (\ref{pm1}) \ depends on certain integration constants
identified with respective physical constants, the target metrics (\ref%
{newbh}) are determined by generating and integrating data, in general,
depending on all space coordinates. The target d-metric involves, in
general, 6 degrees of freedom expressed by functions $\psi (r,\theta ),\eta
_{4}(r,\theta ,\varphi ),w_{i}(r,\theta ,\varphi )$ and $n_{i}(r,\theta
,\varphi )$ related by certain differential or integral formulas to
generating sources (\ref{esourcm}). In GR, the Bianchi identities (we note the
components $\nabla _{\alpha }En^{\alpha 4}=0,$ see formulas (\ref{criccdsc})
but for $En_{\beta \gamma }[\nabla ])$) impose 4 non-dinamical constraints
on the remaining 6 degrees of freedom for metrics and allow only 2 dynamical
degrees of freedom. A corresponding perturbative analysis confirm two
polarizations (i.e. 2 independent degrees of freedom) for gravitons, which
are of spin 2. This holds for linearized gravitational equations reduced to
certain relativistic d'Alambert (wave) operators.  Nevertheless, nonlinear
off-diagonal gravitational interactions (studied in this work for
quasi-stationary target configurations) are determined by 6 degrees of
freedom of solutions of type (\ref{newbh}). Extra degrees of freedom can be
used for the nonlinear modelling of nonlinear structure of gravitational vacuum,
with pattern-forming new structures, filament and/or quasicrystal
configurations etc. as we reviewed in \cite{sv11,vbubuianu17,partner02} and
references therein.
%%%

\subsubsection{Small parametric gravitational polarizations of constants for
a prescribed BH horizon}

Considering small parametric decompositions with $\kappa $-linear terms as
in (\ref{nlims2}), we can provide more exact physical interpretations of
off-diagonal solutions (\ref{newbh}). We apply the AFCDM with a small
parameter and construct d-metrics of type (\ref{offdncelepsilon}) determined
by $\chi $-generating functions: 
\begin{eqnarray*}
d\ \widehat{s}^{2} &=&\widehat{g}_{\alpha \beta }(r,\theta ,\varphi ;\psi
(r,\theta ),g_{4}(r,\theta ,\varphi );\ _{e}^{v}\Upsilon (r,\theta ,\varphi
)+\ _{m}^{v}\Upsilon (r,\theta ,\varphi );\Lambda )du^{\alpha }du^{\beta }=
\\
&&e^{\psi _{0}(r,\theta )}(1+\kappa \ ^{\psi }\chi (r,\theta
))[(dx^{1}(r,\theta ))^{2}+(dx^{2}(r,\theta ))^{2}]-
\end{eqnarray*}%
\begin{eqnarray*}
&&\{\frac{4[\partial _{\varphi }(|\zeta _{4}\mathring{g}_{4}|^{1/2})]^{2}}{%
\mathring{g}_{3}|\int d\varphi \lbrack \ (\ _{e}^{v}\Upsilon +\
_{m}^{v}\Upsilon )\ \partial _{3}(\zeta _{4}\mathring{g}_{4})]|}-\kappa
\lbrack \frac{\partial _{\varphi }(\chi _{4}|\zeta _{4}\mathring{g}%
_{4}|^{1/2})}{4\partial _{\varphi }(|\zeta _{4}\mathring{g}_{4}|^{1/2})}-%
\frac{\int d\varphi \{(\ _{e}^{v}\Upsilon +\ _{m}^{v}\Upsilon )\partial
_{\varphi }[(\zeta _{4}\mathring{g}_{4})\chi _{4}]\}}{\int d\varphi \lbrack
(\ _{e}^{v}\Upsilon +\ _{m}^{v}\Upsilon )\partial _{\varphi }(\zeta _{4}%
\mathring{g}_{4})]}]\}\mathring{g}_{3} \\
&&\{d\varphi +[\frac{\partial _{i}\ \int d\varphi (\ _{e}^{v}\Upsilon +\
_{m}^{v}\Upsilon )\ \partial _{\varphi }\zeta _{4}}{(\breve{N}_{i}^{3})\ (\
_{e}^{v}\Upsilon +\ _{m}^{v}\Upsilon )\partial _{\varphi }\zeta _{4}}+\kappa
(\frac{\partial _{i}[\int d\varphi \ (\ _{e}^{v}\Upsilon +\ _{m}^{v}\Upsilon
)\ \partial _{\varphi }(\zeta _{4}\chi _{4})]}{\partial _{i}\ [\int d\varphi
\ (\ _{e}^{v}\Upsilon +\ _{m}^{v}\Upsilon )\partial _{\varphi }\zeta _{4}]}-%
\frac{\partial _{\varphi }(\zeta _{4}\chi _{4})}{\partial _{\varphi }\zeta
_{4}})]\mathring{N}_{i}^{3}dx^{i}\}^{2}+
\end{eqnarray*}%
\begin{eqnarray}
&&\zeta _{4}(1+\kappa \ \chi _{4})\ \mathring{g}_{4}\{dt+[(\mathring{N}%
_{k}^{4})^{-1}[\ _{1}n_{k}+16\ _{2}n_{k}\int d\varphi \frac{\left( \partial
_{\varphi }[(\zeta _{4}\mathring{g}_{4})^{-1/4}]\right) ^{2}}{|\int d\varphi
\partial _{\varphi }[(\ _{e}^{v}\Upsilon +\ _{m}^{v}\Upsilon )(\zeta _{4}%
\mathring{g}_{4})]|}]+  \label{offdnceleps1} \\
&&\kappa \frac{16\ _{2}n_{k}\int d\varphi \frac{\left( \partial _{\varphi
}[(\zeta _{4}\mathring{g}_{4})^{-1/4}]\right) ^{2}}{|\int d\varphi \partial
_{\varphi }[(\ _{e}^{v}\Upsilon +\ _{m}^{v}\Upsilon )(\zeta _{4}\mathring{g}%
_{4})]|}(\frac{\partial _{\varphi }[(\zeta _{4}\mathring{g}_{4})^{-1/4}\chi
_{4})]}{2\partial _{\varphi }[(\zeta _{4}\mathring{g}_{4})^{-1/4}]}+\frac{%
\int d\varphi \partial _{\varphi }[(\ _{e}^{v}\Upsilon +\ _{m}^{v}\Upsilon
)(\zeta _{4}\chi _{4}\mathring{g}_{4})]}{\int d\varphi \partial _{\varphi
}[\ (\ _{e}^{v}\Upsilon +\ _{m}^{v}\Upsilon )(\zeta _{4}\mathring{g}_{4})]})+%
}{\ _{1}n_{k}+16\ _{2}n_{k}[\int d\varphi \frac{\left( \partial _{\varphi
}[(\zeta _{4}\mathring{g}_{4})^{-1/4}]\right) ^{2}}{|\int d\varphi \partial
_{\varphi }[\ (\ _{e}^{v}\Upsilon +\ _{m}^{v}\Upsilon )(\zeta _{4}\mathring{g%
}_{4})]|}]}]\mathring{N}_{k}^{4}dx^{k}\}^{2}.  \notag
\end{eqnarray}%
The polarization functions $\zeta _{4}(r,\theta ,\varphi )$ and $\chi
_{4}(r,\theta ,\varphi )$ in (\ref{offdnceleps1}) can be prescribed to be of
a necessary smooth class. Such a d-metric describes small $\kappa $%
-parametric off-diagonal deformations of a prime BH d-metric (\ref{pm1a})
when the coefficients became anisotropic on the $\varphi $-coordinate.

If we prescribe a spherical horizon $r=\mathring{h}=2\mathring{m}$ as in (%
\ref{metricf}) with prime data (\ref{pm1a}), we can chose certain smooth
functions for deformations and integration data in (\ref{offdnceleps1}) that
the off-diagonal configurations inside a BH would have a similar (but with a
more rich structure of the gravitational vacuum) as in Section II of \cite%
{casadio25}.

\subsection{Voids and solitons inside of black ellipsoids}

We can generate ellipsoidal deformations of a horizon on $\varphi $ in (\ref%
{offdnceleps1}) if we choose 
\begin{equation}
\chi _{4}(r,\theta ,\varphi )=\underline{\chi }(r,\theta )\sin (\omega
_{0}\varphi +\varphi _{0}).  \label{rotoid}
\end{equation}%
In this formula, $\underline{\chi }(r,\varphi )$ is a smooth function and $%
\omega _{0}$ and $\theta _{0}$ are some constants. For such generating
polarization functions and $\zeta _{4}(r,\theta ,\varphi )\neq 0,$ we obtain
that $(1+\kappa \ \chi _{4})\ \mathring{g}_{4}\simeq (1+\kappa \ \chi _{4})\
f^{+}\simeq 0.$ So, we can approximate $r\simeq 2\mathring{m}/(1+\kappa \
\chi _{4}),$ which is a parametric equation for a rotoid configuration. The
parameter $\kappa $ can be treated as a small eccentricity parameter if the
generating function (\ref{rotoid}) is considered for a $\underline{\chi }%
(r,\theta )\simeq \underline{\chi }=const$. %%%%%

We can prescribe polarization functions generating Schwarzschild BHs embedded
into a nontrivial nonholonomic quasi-stationary background for GR. For small
ellipsoidal deformations of type (\ref{rotoid}), we model black ellipsoid,
BE, objects as generic off-diagonal solutions of the Einstein equations.
Such configurations can be stable, see details and references in \cite%
{vi17,partner06} \ (for MGTs with possible restrictions to GR) if we impose
necessary types of nonholonomic constraints. Selecting generating and
integration functions of type (\ref{zerot1}), we extract LC-configurations
with scalar curvature $R(r,\varphi ,\theta )\simeq \tilde{\Lambda}(r,\varphi
,\theta ),$ in a generalized form as in \cite{ovalle21}. Here we shall
consider also the nonlinear symmetries (\ref{nlims2}) and treat $\tilde{%
\Lambda}(r,\varphi ,\theta )$ as a nonlinear polarization of $\Lambda $.
This modifies (with new terms proportional to $\kappa $) the BH constants
for a effective mass $m(r,\varphi ,\theta )$ and cosmological constant $%
\Lambda .$ Such effects are with local anisotropic polarizations of the
vacuum gravitational background. The phenomenon of warped curvature \cite%
{ovalle21} can be preserved for some subclasses of nonholonomic deformations, 
but the gravitational vacuum is with additional polarizations, and the matter
tensor(\ref{msourc}) transforms into an effective source $Y_{\alpha \beta
}(r,\theta ,\varphi )$ (\ref{esourcm}).

Let us analyze other alternatives of quasi-stationary configurations in
interior of a BH. We consider such a variant: The h-part of the d-metric (%
\ref{newbh}) is prescribed to satisfy instead of a 2-d Poisson equation (\ref%
{poisson}) the generalized Taubes equation for vortices on a curved
background 2-d surface, 
\begin{equation}
_{h}\nabla ^{2}\psi =\Omega _{0}(C_{0}-C_{1}e^{2\psi }),  \label{taubeq}
\end{equation}%
see details in \cite{manton17}. In (\ref{taubeq}), the position-dependent
confrormal factor $\Omega _{0}$ and effective source $(C_{0}-C_{1}e^{2\psi })$
are prescribed as respective generating h-function $\psi (x^{k})$ and
generating h-source $\ ^{h}\Upsilon (x^{k})=\ _{e}^{h}\Upsilon (x^{k})+\
_{m}^{h}\Upsilon (x^{k})$ from (\ref{esourcm}). By re-scaling, both
constants $C_{0}$ and $C_{1}$ take standard values $-1,0,$ or 1 (there are
only five combinations of these values allow vortex solutions $\psi \lbrack
vortex]$ without singularities as was stated for such equations).

The v-part of (\ref{newbh}) can be modelled in a solitonic angular $\varphi $
wave distributions (in the interior of a BH, the time-like coordinate can change
into a space one, and inversely) form for $\eta _{4}(r,\theta ,\varphi )=%
\underline{\eta },$ when 
\begin{equation}
\underline{\eta }\ \simeq \left\{ 
\begin{array}{ccc}
_{r}^{sol}\underline{\eta }(r,\varphi ) & 
\mbox{ as a solution of the modified KdV
equation }\frac{\partial \underline{\eta }}{\partial \varphi }-6\underline{%
\eta }^{2}\frac{\partial \underline{\eta }}{\partial r}+\frac{\partial ^{3}%
\underline{\eta }}{\partial r^{3}}=0, & \mbox{  radial solitons}; \\ 
_{\theta }^{sol}\underline{\eta }(\theta ,\varphi ) & 
\mbox{ as a solution of the
modified KdV equation }\frac{\partial \underline{\eta }}{\partial \varphi }-6%
\underline{\eta }^{2}\frac{\partial \underline{\eta }}{\partial \theta }+%
\frac{\partial ^{3}\underline{\eta }}{\partial \theta \,^{3}}=0, & 
\mbox{ angular
solitons}.%
\end{array}%
\right.  \label{solitonw}
\end{equation}%
We cite \cite{doikou20} and reference therein on such types of solitonic
wave equations which can be transformed into solitonic quasi-stationary
distributions. Invoving a small parameter, we can encode solitonic data in
a $\kappa $-parametric form $\chi _{4}(r,\theta ,\varphi )=\underline{\eta }$
for d-metrics (\ref{offdnceleps1}).

In a series of our and co-authors' works,  we constructed more general
classes of generic off-diagonal cosmological and quasi-stationary solutions
with 3-d solitonic waves and solitonic hierarchies in GR and MGTs, see
reviews and references in \cite{vi17,partner06}. Such configurations can be
used for modelling DM quasi-periodic and pattern-forming structures. In this
subsection, those results were modified in such a form that analogous cosmological solutions are constructed in the framework of the Einstein
gravity theory with respective dualization to quasi-stationary
configurations.
%%%%

\section{Regular BHs and their off-diagonal deformations}

\label{sec4}In this section, we show how regular BHs with Schwarzschild
exterior can be off-diagonally generalized to quasi-stationary solutions of
nonholonomically modified Einstein equations (\ref{cdeq1}) or (\ref{cdeq1a}).

\subsection{Primary metrics and physical constants for regular BHs with
Schwarzshild exterior}

For a primary BH metric (\ref{pm1}), the Ricci scalar for $\mathring{\nabla}$
is computed, 
\begin{equation*}
\mathring{R}=2(rm^{\bullet \bullet }+2m^{\bullet })/r^{2},\mbox{ for }%
0<r\leq \mathring{h}.
\end{equation*}%
Regular BH solutions with $\mathring{R}=\sum\nolimits_{n=2}^{\infty
}c_{n}r^{n-2},$ for $n=2,3,...,$ in vicinity of $r=0,$ are generated by mass
functions (\ref{massf}) of type 
\begin{equation}
m(r)=m_{0}-\frac{q^{2}}{2r}+\frac{1}{2}\sum\nolimits_{n=2}^{\infty }\frac{%
c_{n}r^{n+1}}{(n+1)(n+2)},\mbox{ for }0<r\leq \mathring{h}.  \label{massfreg}
\end{equation}%
According to \cite{ovalle24,casadio25}, the integrations constants $m_{0},$
can be identified with the ADM mass, and $q,$ can be identified with the
electric charge for the Reisner-Nordstr\"{o}m geometry \cite%
{hawking73,misner73,wald82,kramer03}, must be fixed to zero if we want to
have zero values for the Ricci scalar, $\mathring{R}=0,$ the Ricci scalar, $%
\mathring{R}_{\alpha \beta }\mathring{R}^{\alpha \beta }=0,$ and the
Kretschmann scalar, $\mathring{R}_{\alpha \beta \gamma \delta }\mathring{R}%
^{\alpha \beta \gamma \delta }=0,$ that are regular around $r=0.$ We
generate solutions of the standard Einstein equations with energy-momentum
tensor (\ref{emtensint}) if
\begin{equation*}
\varkappa p_{\theta }=-\frac{1}{2}\sum\nolimits_{n=2}^{\infty }\frac{%
nc_{n}r^{n-2}}{n+2}\mbox{ and },\varkappa \epsilon (r)=-\varkappa
p_{r}(r)=\sum\nolimits_{n=2}^{\infty }\frac{c_{n}r^{n-2}}{n+2},\mbox{ for }%
0<r\leq \mathring{h}.
\end{equation*}

In next subsections, we analyze three such regular BH configurations and their quasi-stationary off-diagonal generalizations.

\subsection{Off-diagonal deformations of regular Schwarzchild BHs}

We use a family of prime metrics (\ref{pm1}) with a mass function (\ref{massfreg}), subjected to the continuity conditions (\ref{continucond}) and when the coefficients $c_{n}$ are such way chosen that 
\begin{equation}
_{e}^{\Lambda }m(r)=\frac{r}{2(n-2)}\left( \frac{r}{\mathring{h}}\right)
^{2}[(n+1)-3\left( \frac{r}{\mathring{h}}\right) ^{n-2}]\mbox{ for }n=3,4,...
\label{massf1}
\end{equation}%
The left labels $e$ and $\Lambda $ are used because such a configuration is with an effective cosmological constant as we shall explain below. The integer  $n$ labels a family of regular BHs (it is not a hair).  The mass function $_{e}^{\Lambda }m(r)$ (\ref{massfreg})  results in respective $n$%
-parametriezed $f^{-}$ in (\ref{metricf}), $p_{\theta }(r)$ and $\epsilon (r)=-p_{r}(r)$ considered above and (\ref{emtensint}). Such expressions are valid for $0<r\leq \mathring{h};$ define fluid-like sources for the BH solutions; satisfy the weak energy condition; and represent alternative
sources for the Schwarzschild exterior. The energy density and pressures behave monotonically as stated in section IIIA of \cite{casadio25}, see also details in \cite{ovalle24}.
%%%%%

We emphasize that $_{e}^{\Lambda }m(r)$ (\ref{massfreg}) results in a
property that in the region near $r=0,$ we obtain a de Sitter like behaviour
with effective cosmological constant $^{e}\Lambda =3/\mathring{h}.$ This
results in corresponding behaviours: 
\begin{equation*}
\ _{e}^{\Lambda }f(r)\sim 1-\frac{r^{2}}{\mathring{h}^{2}}\mbox{ and }\
_{e}^{\Lambda }\epsilon (r)=-\ _{e}^{\Lambda }p_{r}(r)\sim -\ _{e}^{\Lambda
}p_{\theta }(r)\sim \frac{3}{\varkappa \mathring{h}^{2}}\sim \frac{\
_{e}^{\Lambda }\mathring{R}}{4}.
\end{equation*}%
For increasing $n,$ the de Sitter core grows towards the horizon and statistically maches the Schwarzschild exterior. Here we note also that in addition to the horizon $\mathring{h}=2\mathring{m}$ the prime solutions possess also a Cauchy horizon $h_{c}=\sqrt{2/3}\mathring{h}$ as stated in Fig 1 of \cite{casadio25}.

Using the techniques described in Appendix \ref{appendixa}, we can perform
off-diagonal quasi-stationary generalizations of prime metrics (\ref{pm1})
and, respective, prime d-metrics (\ref{pm1a}), when $m(r)\rightarrow $ $%
_{e}^{\Lambda }m(r)$ (\ref{massf1}) and $f(r)\rightarrow \ _{e}^{\Lambda
}f(r)$ as described above. For regular configurations, $(\mathring{g}%
_{\alpha },\mathring{N}_{i}^{a})\rightarrow (\ _{e}^{\Lambda }\mathring{g}%
_{\alpha },\ _{e}^{\Lambda }\mathring{N}_{i}^{a})$ and 
\begin{equation*}
\ _{e}\widehat{\mathbf{\Upsilon }}_{\ \ \beta }^{\alpha }=diag[\
_{e}^{h}\Upsilon ,\ _{e}^{v}\Upsilon ]\rightarrow \ _{e}^{\Lambda }\widehat{%
\mathbf{\Upsilon }}_{\ \ \beta }^{\alpha }=diag[\ _{e}^{h}\Upsilon (\
_{e}^{\Lambda }m(r)),\ _{e}^{v}\Upsilon (\ _{e}^{\Lambda }m(r))],
\end{equation*}%
respectively in formulas (\ref{msourc}) and (\ref{esourcm}). Nonlinear
symmetries (\ref{nlims2}) can be adapted to the condition $\Lambda =\
^{e}\Lambda .$ Introducing such transforms of primary data into respective
classes of quasi-stationary solutions (\ref{newbh}) or (\ref{offdnceleps1}),
we define regular off-diagonal modifications of the Schwarzschild exterior.
For generating functions (\ref{rotoid}), we generate ellipsoidal exterior
configurations.

\subsection{Distorted regular BHs with vanishing angular pressure}

The off-diagonal modifications can be performed in a different way for the
case of regular Schwarzschild BH with $p_{\theta }(\mathring{h})=0.$As we
mentioned for the continuity conditions (\ref{continucond}), $p_{\theta }(r) 
$ can be discontinuous across $r=\mathring{h}.$We can generate a family of
prime metrics (\ref{pm1}) labeled by two integer constants $n$ and $l$ if we
impose the conditions $p_{\theta }(\mathring{h})=0$ and $m^{\bullet \bullet
}(\mathring{h})=0.$ An example by such mass function (\ref{massfreg}) is
given by 
\begin{equation}
_{e}^{l}m(r)=\frac{(n+1)(l+1)r}{2(n-2)(l-2)}\left( \frac{r}{\mathring{h}}%
\right) ^{2}[1-\frac{3(l-2)}{(n+1)(l-n)}\left( \frac{r}{\mathring{h}}\right)
^{n-2}+\frac{3(n-2)}{(l+1)(l-n)}\left( \frac{r}{\mathring{h}}\right) ^{l-2}]
\label{massf2}
\end{equation}%
for integer $2<n<l=4,5,...$ We use left labels $l$ and $e$ emphasize that
this effective mass is determined additionally by an integer $l,$ which is
different from $_{e}^{\Lambda }m(r)$ (\ref{massf1}).

We modify respectively (\ref{pm1}) and (\ref{pm1a}), when $m(r)\rightarrow $ 
$_{e}^{l}m(r)$ (\ref{massf2}) and $f(r)\rightarrow \ _{e}^{l}f(r).$ For
regular configurations, $(\mathring{g}_{\alpha },\mathring{N}%
_{i}^{a})\rightarrow (\ _{e}^{l}\mathring{g}_{\alpha },  \ _{e}^{l}\mathring{%
N}_{i}^{a})$ and 
\begin{equation*}
\ _{e}\widehat{\mathbf{\Upsilon }}_{\ \ \beta }^{\alpha }=diag[\
_{e}^{h}\Upsilon ,\ _{e}^{v}\Upsilon ]\rightarrow \ \ _{e}^{l}\widehat{%
\mathbf{\Upsilon }}_{\ \ \beta }^{\alpha }=diag[\ _{e}^{h}\Upsilon (\ \
_{e}^{l}m(r)),\ _{e}^{v}\Upsilon (\ \ _{e}^{l}m(r))],
\end{equation*}%
are used respectively in formulas (\ref{msourc}) and (\ref{esourcm}).  The
two integer parameterized family of such prime BH solutions results are 
determined by a respective family of matter sources $ \ _{e}^{l}\epsilon
(r)=-\ _{e}^{l}p_{r}(r)$ and $-\ _{e}^{l}p_{\theta }(r).$ Such conditions
define a simplest regular BH with Schwarzschild exterior,  the
energy-momentum tensor is continuous across the horizon (i.e.$\ \ _{e}^{l}%
\widehat{\mathbf{T}}_{\ \ \beta }^{\alpha }(\mathring{h})=0$). For such
configuration, we do not have an induced effective $\ ^{e}\Lambda .$
Neverthelee, for off-diagonal generalizations of (\ref{pm1}) and (\ref{pm1a}%
), we can introduce a nontrivial $\Lambda ,$ which via nonlinear symmetries (%
\ref{nlims2}) characterize off-diagonal contributions in the modified
Schwarzschild exterior. An example with $n=3$ and $l=4$ in section IIIB of 
\cite{casadio25} show that respective mass functions define respective Cauchy
inner horizons at $r=\mathring{h}/2.$

We note that introducing primary data $(\ _{e}^{l}\mathring{g}_{\alpha },\ \
_{e}^{l}\mathring{N}_{i}^{a})$ and above $\ _{e}^{l}\widehat{\mathbf{%
\Upsilon }}_{\ \ \beta }^{\alpha }$ into respective classes of target
quasi-stationary solutions (\ref{newbh}) or (\ref{offdnceleps1}), we define
regular off-diagonal and $(n,l)$ - parameterized modifications of the
Schwarzschild exterior. For generating functions (\ref{rotoid}), we generate
ellipsoidal exterior configurations. Such regular off-diagonal deformations
may encode additional solitonic configurations, some void structures,
pattern-forming data, quasi-periodic structures, etc., with quasi-periodic $%
(n,l)$-labels.

Finally, we emphasize that using the AFCDM we can study off-diagonal
generalizations for the extremal Schwarzschild BH studied in section IIIC of 
\cite{casadio25}; and dual cosmological metrics (examples of Kantowski-Sachs
homogeneous by anisotropic universes) studied in section IV of that work.
Such constructions were performed in our former works, in the main for MGTs,
see \cite{vi17,partner06} and references therein. The goal of this work is
to concentrate on regular, in general, off-diagonal quasi-stationary
deformations of BH solutions parameterized by certain integers $n$ and $l.$

\section{G. Perelman's thermodynamics for relativistic geometric flows}

\label{sec5}In general, the off-diagonal deformations of the Schwarzschild BH metrics constructed in the previous two sections are not characterized by certain hypersurface or holographic conditions. The Bekenstein--Hawking BH thermodynamic paradigm \cite{bek2,haw2} is applicable, for instance, for BE configurations (\ref{rotoid}) when the hyperurface area can be defined and used for definition of the BE entropy and temperature. It is similar to the case of Kerr BHs but with generic off-diagonal terms of rotoid configuration.
%%%%%

For general off-diagonal deformations in GR (for two different such classes of quasi-stationary or locally anisotropic cosmological solutions), the mentioned thermodynamic approach is not applicable. To characterize physical properties of new classes of exact or parametric quasi-stationary solutions, we proposed to apply G. Perelman's approach to statistical thermodynamics of Ricci flows \cite{perelman1} by generalizing the constructions in relativistic and nonholonomic forms \cite{svnonh08,gheorghiuap16,partner06}, see references therein.  
%%%%%

In our works on generalizations and applications of geometric flow theories, we do not attempt to formulate/ prove any modified variant of famous Poincar\'{e}-Thurston conjecture are not formulated/ proved. For relativistic geometric flows, this is a very difficult mathematical problem, which became
undetermined in the case of various nonassociative/ noncommutative, or supersymmetric, nonmetric, etc. theories (for instance, because of an infinite number of noncommutative differential and integral calculi which can be formulated). Here we note that mathematical rigorous proofs of the results for Ricci flows of Riemannian metrics consist of some hundreds of pages as in monographs \cite{monogrrf1,monogrrf2,monogrrf3}. So, it is not a goal of this work to develop those topological and geometric analysis methods for Ricci flows, including the relativistic Einstein equations as a particular case, i.e. as a relativistic Ricci solitons. Nevertheless,  some sections of \cite{perelman1} related to the concepts of G. Perelman's W-entropy and derived statistical/ geometric thermodynamics can be
generalized in a straightforward form to relativistic flows and applied to characterizing physical properties of all classes of solutions, for instance, in GR. In a certain sence, G. Perelman's thermodynamic paradigm is an alternative for studying off-diagonal solutions which can't be considered in the framework of Bekenstein-Hawking thermodynamics of BHs.
%%%%%

In Appendix \ref{appendixb}, we summarize necessary results on relativistic Ricci flows encoding geometric evolution of nonholonomic Einstein systems (NES). Relativistic generalizations of the R. Hamilton \cite{hamilton82} and D. Friedan \cite{friedan80} geometric flow equations\footnote{in mathematical literature, they are known as R. Hamilton's equations but physicists considered independently and few years early some equivalents of such equations} are postulated in such forms that they describe in self-consistent form the evolution of large classes of off-diagonal solutions for NES. This is different from \cite{papad24} when the Einstein flows are not used but only Ricci flows (which is more convenient for investigating sigma-models in supersymmetric theories and related c-theorem
issue). The main goals of this section is to formulate a relativistic variant of the G. Perelan thermodynamics and show how respective thermodynamic variables can be computed for off-diagonal deformations of the regular BH solutions constructed in sections \ref{sec3} and \ref{sec4}.
%%%%%

\subsection{Nonholonomic flows of geometric objects in GR}

In the theory of (relativistic) Ricci flows, $\tau $-families of metrics $g_{\alpha \beta }(\tau ):=g_{\alpha \beta }(\tau ,u^{\gamma })$ are considered, where $\tau $ is a positive flow parameter (treated as a conventional temperature according to \cite{perelman1}). In brief, we shall write $g_{\alpha \beta }(\tau ),\nabla (\tau ),\mathbf{D}(\tau ),\widehat{\mathbf{R}}ic(\tau ),...$ if that will not result in ambiguities (the coordinate dependence will be stated explicitly when necessary). The geometric $\tau $-evolution of fundamental geometric and physical geometric d-objects is defined by (nonholonomic) geometric flow evolution equations as
described in Appendix \ref{appendixb}. In our approach, we formulate them in canonical (hat) nonholonomic variables, which allows us to apply the AFCDM
and decouple and solve the relativistic versions of R. Hamilton's equations in certain parametric forms. The nonholonomic variables and normalizing
functions are introduced in such forms that for a fixed $\tau =\tau _{0},$ the nonholonomic Ricci flow equations define self-similar configurations,
i.e. nonholonomic Ricci solitons which for certain conditions are equivalent to the nonholonomic Einstein equations (\ref{einstceq1}) and (\ref{cdeq1}),
or (\ref{cdeq1a}). The nonholonomic canonical dyadic structures are N-adapted in such forms that all geometric d-objects; dynamical gravitational equations and their off-diagonal solutions; constraints and nonlinear symmetries are describes by an additional dependence on $\tau ,$ at least for an interval $0<$ $\tau _{1}<$ $\tau <$ $\tau _{2}.$
%%%%

\subsection{Relativistic W-entropy and statistical and geometric thermodynamics}

In this subsection, we generalize G. Perelman's thermodynamics \cite{perelman1} for the Einstein gravity theory. To perform the constructions in canonical dyadic variables, we perform a change of the normalization function for the functionals (\ref{fperelmNES}) and (\ref{wfperelmNES}), 
$\zeta (\tau)\rightarrow \widehat{\zeta }(\tau ).$ Let us our purposes, we consider a $\widehat{\zeta }(\tau )$ defined by 
\begin{equation*}
\partial _{\tau }\zeta (\tau )+\widehat{\square }(\tau )[\zeta (\tau
)]-\left\vert \widehat{\mathbf{D}}\zeta (\tau )\right\vert ^{2}-\ \widehat{%
\mathcal{L}}(\tau )=\partial _{\tau }\ \widehat{\zeta }(\tau )+\widehat{%
\square }(\tau )[\widehat{\zeta }(\tau )]-\left\vert \widehat{\mathbf{D}}%
\widehat{\zeta }(\tau )\right\vert ^{2}.
\end{equation*}%
Such equations can be solved by a factorization of variables. This is always possible if we use (\ref{normcond}) and a $\widehat{\zeta }(\tau )$ as a
solution of 
\begin{equation}
\partial _{\tau }\widehat{\zeta }(\tau )=-\widehat{\square }(\tau )[\widehat{%
\zeta }(\tau )]+\left\vert \widehat{\mathbf{D}}\widehat{\zeta }(\tau
)\right\vert ^{2}-\widehat{\mathbf{R}}sc(\tau ).  \label{normcondc}
\end{equation}%
Two normalization functions, $\zeta (\tau )$ or $\widehat{\zeta }(\tau ),$  define different integration measures in topological type theories. In our approach, we use different types of transforms $\zeta (\tau )\rightarrow \widehat{\zeta }(\tau )$ which allow us to absorb, or (inversely) distinguish different types of (effective) $\tau $-running Lagrange densities. Such transforms may simplify the formulas and the procedure of finding exact and parametric solutions. 
%%%%%

In terms of an integration measure with a $\widehat{\zeta }(\tau )$ (\ref{normcondc}), the W-functional (\ref{wfperelmNES}) can be written in the form 
\begin{equation}
\ \widehat{\mathcal{W}}(\tau )=\int_{t_{1}}^{t_{2}}\int_{\Xi _{t}}\left(
4\pi \tau \right) ^{-2}e^{-\widehat{\zeta }(\tau )}\sqrt{|\ \mathbf{g}(\tau
)|}\delta ^{4}u[\tau (\widehat{\mathbf{R}}sc(\tau )+|\widehat{\mathbf{D}}%
(\tau )\widehat{\zeta }(\tau )|^{2}+\widehat{\zeta }(\tau )-4].  \label{wf1}
\end{equation}%
In this formula, the effective nonholonomic and matter fields sources $\widehat{\mathcal{L}}(\tau )$ can be encoded into geometric data. We have to consider an explicit class of solutions of (\ref{ricciflowr2}) related for $\tau =\tau _{0}$ to solutions of the NES equations (\ref{cdeq1}), or (\ref{canriccisolda}), and extended to (\ref{ricciflowr2}). In the original version for 3-d LC-configurations \cite{perelman1}, the  functional (\ref{wf1}) is called W-entropy because it has the properties of "minus entropy".
%%%%%%

Using double nonholonomic splitting 2+2 and 3+1 functionals on a nonholonomic Lorentz manifold $\mathbf{V,}$ as we explained in Appendix \ref%
{appendixb} for (\ref{fperelmNES}) and (\ref{wfperelmNES}), we can introduce such a statistical partition function: 
\begin{equation}
\ \widehat{Z}(\tau )=\exp [{\int_{\widehat{\Xi }}[-\widehat{\zeta }+2]\
\left( 4\pi \tau \right) ^{-2}e^{-\widehat{\zeta }}\ \delta \widehat{%
\mathcal{V}}(\tau )}],  \label{spf}
\end{equation}%
where the volume element $\delta \widehat{V}(\tau )$ is defined and computed for any $\tau $-families of d-metrics (\ref{dm}), 
\begin{equation}
\delta \widehat{V}(\tau )=\sqrt{|\mathbf{g}(\tau )|}\ dx^{1}dx^{2}\delta
y^{3}\delta y^{4}\ .  \label{volume}
\end{equation}%
%%%%%%

Using $\widehat{Z}(\tau )$ (\ref{spf}), we can generalize the variational procedure provided in section 5 of \cite{perelman1}) by using a N-adapted variational procedure related to $\widehat{\mathcal{W}}(\tau )$ (\ref{wf1}) (considered on a closed region in $\mathbf{V)}.$ Alternatively, we can use an abstract nonholonomic geometric approach with respective distortions. In both cases, we can define and compute such canonical thermodynamic variables: 
\begin{align}
\ \widehat{\mathcal{E}}\ (\tau )& =-\tau ^{2}\int_{\widehat{\Xi }}\ \left(
4\pi \tau \right) ^{-2}\left( \widehat{\mathbf{R}}sc+|\ \widehat{\mathbf{D}}%
\ \widehat{\zeta }|^{2}-\frac{2}{\tau }\right) e^{-\widehat{\zeta }}\ \delta 
\widehat{V}(\tau ),  \label{qthermvar} \\
\ \ \ \widehat{S}(\tau )& =-\ \widehat{\mathcal{W}}(\tau )=-\int_{\widehat{%
\Xi }}\left( 4\pi \tau \right) ^{-2}\left( \tau (\widehat{\mathbf{R}}sc+|%
\widehat{\mathbf{D}}\widehat{\zeta }|^{2})+\widehat{\zeta }-4\right) e^{-%
\widehat{\zeta }}\ \delta \widehat{V}(\tau ),  \notag \\
\ \ \ \widehat{\sigma }(\tau )& =2\ \tau ^{4}\int_{\widehat{\Xi }}\left(
4\pi \tau \right) ^{-2}|\ \widehat{\mathbf{R}}_{\alpha \beta }+\widehat{%
\mathbf{D}}_{\alpha }\ \widehat{\mathbf{D}}_{\beta }\widehat{\zeta }_{[1]}-%
\frac{1}{2\tau }\mathbf{g}_{\alpha \beta }|^{2}e^{-\widehat{\zeta }}\ \delta 
\widehat{V}(\tau ).  \notag
\end{align}%
Re-defining the normalizing functions and adapting the nonholonomic distributions, these variables can be formulated for LC configurations (by imposing constraints (\ref{lccond1}) or (\ref{lccond}). Here we note that the quadratic fluctuation thermodynamic variable $\widehat{\sigma }(\tau )$ can be written as a functional of $\widehat{\mathbf{R}}_{\alpha \beta }$ when $\widehat{\mathcal{E}}\ (\tau )$ and $\widehat{S}(\tau )$ are functionals of $\widehat{\mathbf{R}}sc.$ In this work, we omit cumbersome technical details on explicit computing $\widehat{\sigma }(\tau )$ for certain classes of off-diagonal solutions.
%%%%

Finally, in this subsection, we conclude that the modified G. Perelman thermodynamic variables (\ref{qthermvar}) characterizing the relativistic geometric flow evolution of NESs can be computed in explicit form for any solution of systems of nonlinear PDEs of type (\ref{cdeq1}), or (\ref{canriccisolda}), and (\ref{ricciflowr2}). The procedure of computing such physical values in explicit form is simplified substantially if we consider solutions with nonlinear symmetries and $\tau $-families of generating functions $\Phi (\tau )$ resulting in $\tau $-evolution of nonholonomic Einstein equations of type (\ref{cdeq1a}). For such models, we assume a $\tau $-running of effective cosmological constants, $\Lambda (\tau ).$ Such a behaviour can be determined in a form to be compatible with observational data for DE or certain DM quasi-stationary distributions.
%%%

\subsubsection{Explicit formulas for thermodynamic variables for quasi-stationary NESs}

We consider $\tau $-families of quasi-stationary d-metrics $\mathbf{g}_{\alpha }[\Phi (\tau )]\simeq \ \mathbf{g}_{\alpha }[\eta _{4}(\tau )]$ 
(\ref{offdiagcosmcsh}) with possible re-definition of generating data subjected to nonlinear symmetries (\ref{nonlintrsmalp}) and (\ref{nonlinsymrex}). The generating sources $\widehat{\mathbf{\Upsilon }}_{\ \ \beta }^{\alpha }(\tau )=[\ ^{h}\Upsilon (\tau ),\ ^{v}\Upsilon (\tau )]$ (\ref{esourc}) are respectively substituted by $\widehat{\mathbf{J}}(\tau)=[\ ^{h}\widehat{\mathbf{J}}(\tau ),\ ^{v}\widehat{\mathbf{J}}(\tau )]$ (\ref{effrfs}); the gravitational polarization is of type $\eta _{4}(\tau )=\eta _{4}(\tau ,x^{i},y^{3});$ and the cosmological constant $\Lambda $ is changed into a $\tau $-running one, $\Lambda (\tau ).$ Such assumptions allow us to express the nonholonomic geometric flow equations (\ref{ricciflowr2}) as a $\tau $-family of nonholonomic Einstein equations 
(\ref{cdeq1a}), $\widehat{\mathbf{R}}_{\ \ \beta }^{\alpha }[\Phi (\tau ),\widehat{\mathbf{J}}(\tau )]=\Lambda (\tau )\mathbf{\delta }_{\ \ \beta}^{\alpha }.$

Using $\widehat{\mathbf{R}}sc(\tau )=4\Lambda (\tau ),$ we can simplify the computation of $\widehat{\mathcal{E}}(\tau )$ and $\widehat{\mathcal{S}}%
(\tau )$ from (\ref{qthermvar}) by introducing a conventional statistical partition function $\ ^{q}\widehat{Z}(\tau )$ for computing the
thermodynamic variables: 
\begin{eqnarray}
\ ^{q}\widehat{Z}(\tau ) &=&\exp [\int_{\widehat{\Xi }}\frac{1}{8\left( \pi
\tau \right) ^{2}}\ \delta \ ^{q}\mathcal{V}(\tau )],\ ^{q}\widehat{\mathcal{%
E}}\ (\tau )=-\tau ^{2}\int_{\widehat{\Xi }}\ \frac{1}{8\left( \pi \tau
\right) ^{2}}[2\Lambda (\tau )-\frac{1}{\tau }]\ \delta \ ^{q}\mathcal{V}%
(\tau ),  \notag \\
\ ^{q}\widehat{S}(\tau ) &=&-\ \ ^{q}\widehat{W}(\tau )=-\int_{\widehat{\Xi }%
}\frac{1}{4\left( \pi \tau \right) ^{2}}[\tau \Lambda (\tau )\ -1]\delta \
^{q}\mathcal{V}(\tau ).  \label{thermvar1}
\end{eqnarray}%
In these formulas, the label $q$ is used for quasi-stationary $\tau $-running configurations. Here we note that the volume elements (\ref{volume}), i.e. 
$\delta \ ^{q}\mathcal{V}(\tau )=\sqrt{|\ ^{q}\mathbf{g}(\tau )|}\ dx^{1}dx^{2}\delta y^{3}\delta y^{4},$ are determined by respective classes
of quasi-stationary solutions. Further computations \ of (\ref{thermvar1}) are simplified for such assumptions: we chose such frame/ coordinate systems
when the normalizing functions have the properties $\widehat{\mathbf{D}}_{\alpha }\ {\widehat{\zeta }}=0$ and ${\widehat{\zeta }}\approx 0;$ we can
consider trivial integration functions $\ _{1}n_{k}=0$ and $\ _{2}n_{k}=0$ (such conditions change for arbitrary frame and coordinate transforms). 
%%%%

\subsubsection{Volume forms for $\protect\eta $-generating or $\chi $-generating functions}

The formulas (\ref{nonlinsymrex}) for nonlinear symmetries allow us to express 
\begin{equation}
\ \Phi (\tau )=2\sqrt{|\ \Lambda (\tau )\ g_{4}(\tau )|}=\ 2\sqrt{|\ \Lambda
(\tau )\ \eta _{4}(\tau )\ \mathring{g}_{4}(\tau )|}\simeq 2\sqrt{|\ \Lambda
(\tau )\ \zeta _{4}(\tau )\ \mathring{g}_{4}|}[1-\frac{\varepsilon }{2}\chi
_{4}(\tau )],  \notag
\end{equation}%
when $\ [\ \Psi (\tau ),\ ^{v}\widehat{\mathbf{J}}(\tau )]\rightarrow
\lbrack \Phi (\tau ),\Lambda (\tau )].$ Then we compute  
\begin{eqnarray*}
\ \delta \ ^{q}\mathcal{V} &=&\delta \mathcal{V}[\tau ,\ \widehat{\mathbf{J}}%
(\tau ),\ \Lambda (\tau );\psi (\tau ),\ g_{4}(\tau )]=\delta \mathcal{V}(\ 
\widehat{\mathbf{J}}(\tau ),\Lambda (\tau ),\eta _{4}(\tau )\ \mathring{g}%
_{4}) \\
&=&\frac{1}{|\ \Lambda (\tau )|}\ \delta \ _{\eta }\mathcal{V},\mbox{ where }%
\ \delta \ _{\eta }\mathcal{V}=\ \delta \ _{\eta }^{h}\mathcal{V}\times
\delta \ _{\eta }^{v}\mathcal{V},\mbox{ for }
\end{eqnarray*}%
\begin{eqnarray}
\delta \ _{\eta }^{h}\mathcal{V} &=&\delta \ _{\eta }^{h}\mathcal{V}[\ ^{h}%
\widehat{\mathbf{J}}(\tau ),\eta _{1}(\tau )\ \mathring{g}_{1}]
\label{volumfuncts} \\
&=&e^{\widetilde{\psi }(\tau )}dx^{1}dx^{2}=\sqrt{|\ ^{h}\widehat{\mathbf{J}}%
(\tau )|}e^{\psi (\tau )}dx^{1}dx^{2},  \notag \\
&&\mbox{ for }\psi (\tau )\mbox{ being a
solution of  }(\ref{poisson})\mbox{ or }(\ref{taubeq}),\mbox{ with sources }%
\ ^{h}\widehat{\mathbf{J}}(\tau );  \notag \\
\delta \ _{\eta }^{v}\mathcal{V} &=&\delta \ _{\eta }^{v}\mathcal{V}[\ ^{v}%
\widehat{\mathbf{J}}(\tau ),\eta _{4}(\tau ),\ \mathring{g}_{4}]  \notag \\
&=&\frac{\partial _{3}|\ \eta _{4}(\tau )\ \mathring{g}_{4}|^{3/2}}{\ \sqrt{%
|\int dy^{3}\ \ _{{}}^{v}\widehat{\mathbf{J}}(\tau )\{\partial _{3}|\ \eta
_{4}(\tau )\ \mathring{g}_{4}|\}^{2}|}}[dy^{3}+\frac{\partial _{i}\left(
\int dy^{3}\ \ ^{v}\widehat{\mathbf{J}}(\tau )\partial _{3}|\ \eta _{4}(\tau
)\ \mathring{g}_{4}|\right) dx^{i}}{\ \ \ ^{v}\widehat{\mathbf{J}}(\tau
)\partial _{3}|\ \eta _{4}(\tau )\mathring{g}_{4}|}]dt.  \notag
\end{eqnarray}%
We can integrate the products of forms from (\ref{volumfuncts}) on a closed hypersurface $\widehat{\Xi }$ and obtain a running spacetime volume
functional 
\begin{equation}
\ _{\eta }^{\mathbf{J}}\mathcal{V[}\ ^{q}\mathbf{g}(\tau )]=\int_{\ \widehat{%
\Xi }}\delta \ _{\eta }\mathcal{V}(\ ^{v}\widehat{\mathbf{J}}(\tau ),\ \eta
_{\alpha }(\tau ),\mathring{g}_{\alpha }).  \label{volumf1}
\end{equation}%
The explicit form of this functional depends on the type of quasi-stationary solutions and geometric data on $\widehat{\Xi }$ and on effective sources for NES. The dependence on primary d-metrics and generating functions $\eta _{4}(\tau )$ and effective cosmological constants $\Lambda (\tau )$ is also involved. The formula (\ref{volumf1}) has a functional character because to compute $\ \ _{\eta }^{\mathbf{J}}\mathcal{V[}\ ^{q}\mathbf{g}(\tau )]$ in a general form for all classes of quasi-stationary solutions is not possible. Nevertheless, it allows us to separate in the thermodynamic variables the terms depending only on $\Lambda (\tau )$ and some coefficients depending only on temperature $\tau $.

Using the volume functional (\ref{volumf1}), we obtain such formulas for the nonholonomic canonical thermodynamic variables (\ref{thermvar1}): 
\begin{eqnarray}
\ _{\eta }^{q}\widehat{Z}(\tau ) &=&\exp \left[ \frac{1}{8\pi ^{2}\tau ^{2}}%
\ \ _{\eta }^{\mathbf{J}}\mathcal{V}[\ ^{q}\mathbf{g}(\tau )]\right] ,\
_{\eta }^{q}\widehat{\mathcal{E}}\ (\tau )=\ \frac{1-2\tau \ \Lambda (\tau )%
}{8\pi ^{2}\tau }\ \ \ _{\eta }^{\mathbf{J}}\mathcal{V[}\ ^{q}\mathbf{g}%
(\tau )],  \label{thermvar2} \\
\ \ \ \ _{\eta }^{q}\widehat{S}(\tau ) &=&-\ _{\chi }^{q}\widehat{W}(\tau )=%
\frac{1-\Lambda (\tau )}{4\pi ^{2}\tau ^{2}}\ \ _{\eta }^{\mathbf{J}}%
\mathcal{V[}\ ^{q}\mathbf{g}(\tau )].  \notag
\end{eqnarray}%
For instance, we can compare a conventional thermodynamic energy $\ ^{q}\widehat{\mathcal{E}}\ (\tau )$ behavior of two nonholonomic quasi-stationary Ricci solitons for $\tau _{1}$ and $\tau _{2}.$ We can speculate also on more or less probable Ricci soliton configurations defined by $\tau $-families of two quasi-stationary metrics $\ _{1}^{q}\mathbf{g}(\tau )$ and $\ _{2}^{q}\mathbf{g}(\tau ),$ when 
\begin{equation}
\ \ _{1}^{q}\widehat{S}(\tau )/\ \ _{2}^{q}\widehat{S}(\tau )=\ _{\eta }^{%
\mathbf{J}}\mathcal{V[}\ _{1}^{q}\mathbf{g}(\tau )]/\ _{\eta }^{\mathbf{J}}%
\mathcal{V[}\ _{1}^{q}\mathbf{g}(\tau )].  \label{fractions}
\end{equation}%
Such relations can be always derived using the nonlinea symmetries (\ref{nonlinsymrex}). $\ $In many cases, we can conclude on thermodynamic and
informational priorities of two different quasi-stationary solutions even to compute in explicit form $\ \ _{\eta }^{\mathbf{J}}\mathcal{V[}\ ^{q}\mathbf{%
g}(\tau )]$ may be a difficult technical task, see details and examples in \cite{partner06}).

Similarly,  nonholonomic Ricci flow thermodynamic variables are defined and computed for quasi-stationary solutions with $\chi $-generating functions resulting in $\kappa $-parametric d-metrics $_{\kappa }^{q}\mathbf{g}(\tau )$ of type (\ref{offdncelepsilon}). In abstract geometric form, we change $\eta \rightarrow
\chi $ in (\ref{thermvar2}) and write 
\begin{eqnarray}
\ _{\chi }^{q}\widehat{Z}(\tau ) &=&\exp \left[ \frac{1}{8\pi ^{2}\tau ^{2}}%
\ \ _{\chi }^{\mathbf{J}}\mathcal{V}[\ ^{q}\mathbf{g}(\tau )]\right] ,\
_{\chi }^{q}\widehat{\mathcal{E}}\ (\tau )=\ \frac{1-2\tau \ \Lambda (\tau )%
}{8\pi ^{2}\tau }\ \ \ _{\chi }^{\mathbf{J}}\mathcal{V[}\ ^{q}\mathbf{g}%
(\tau )],  \label{thermvar3} \\
\ \ \ \ _{\chi }^{q}\widehat{S}(\tau ) &=&-\ _{\chi }^{q}\widehat{W}(\tau )=%
\frac{1-\Lambda (\tau )}{4\pi ^{2}\tau ^{2}}\ \ _{\chi }^{\mathbf{J}}%
\mathcal{V[}\ ^{q}\mathbf{g}(\tau )].  \notag
\end{eqnarray}%
Such thermodynamic variables can be computed for rotoid configurations (\ref{rotoid}). Choosing respective nonholonomic distributions, we can generate
BE configurations (see \cite{vi17} and references therein) which are characterized by (\ref{thermvar3}). Because such $\tau $-families of ellipsoids have respective hypersurface areas $A_{BE}(\tau )$, we can also compute the Bekenstein-Hawking entropy, energy and temperature, denoted respectively $^{BH}S(A_{BE}(\tau )),\ ^{BH}E(A_{BE}(\tau ))$ and $\ ^{BH}temp(A_{BE}(\tau )),$ which for $\tau =\tau _{0}$ are computed as ellipsoid deformations of standard formulas from \cite{hawking73,misner73,wald82}. In such special cases, we can introduce a common calibration for the Hawking temperature and G. Perelman temperature
parameter $\ ^{BH}temp(A_{BE}(\tau ))\sim \tau $ at least of a fixed $\tau =\tau _{0}.$ We can also compare two BE thermodynamic models given by 
$^{BH}S(A_{BE}(\tau ))$ and $\ ^{BH}E(A_{BE}(\tau ))$ and, respectively $\ _{\chi }^{q}\widehat{S}(\tau )$ and 
$\ _{\chi }^{q}\widehat{\mathcal{E}}(\tau)$ from (\ref{thermvar3}). Nevertheless, two thermodynamic descriptions of BEs are not equivalent because the Perelman paradigm is formulated for geometric flow evolution (in this case for quasi-stationary NESs and respective volume forms), but the Bekenstein-Hawking paradigm is rigorously related to certain hypersurface areas.

Finally, we emphasize that for $\tau =\tau _{0},$ the formulas (\ref{thermvar2}) can be used for defining general thermodynamic characteristics of nonholonomic Ricci soliton quasi-stationary configurations, stated by (\ref{canriccisolda}). This can be performed for general off-diagonal solutions even their physical interpretation as BH configurations is not possible.
%%%%%%

\subsection{G. Perelman thermodynamic variables for off-diagonal deformations of regular BHs}

Let us show how to compute in explicit form G. Perelman's thermodynamic variables for physically important off-diagonal solutions constructed in sections \ref{sec3} and \ref{sec4}. In certain special cases of prime metrics or rotoid deformations of the Schwarzschild BHs (for instance, with generating function (\ref{rotoid})), we construct quasi-stationary ellipsoid configurations. In such cases, we can apply the Bekenstein-Hawking thermodynamics \cite{bek2,haw2}. For more general $\tau $-families of off-diagonal deformations of static BHs, we do not have certain well-defined horizons or holographic conditions. Changing the thermodynamic paradigm, we can consider the thermodynamic variables (\ref{thermvar2}). In such cases, we have to compute the respective volume forms 
$\ _{\eta }^{\mathbf{J}}\mathcal{V[}\ ^{q}\mathbf{g}(\tau )]$ (\ref{volumf1}) when $\ ^{q}\mathbf{g}(\tau )$ is defined by corresponding $\tau $-families of quasi-stationary solutions.
%%%%%

\subsubsection{Geometric flow thermodynamics in interior of BHs}

Four general classes of geometric flow thermodynamic models can be constructed for respective types of $\eta $-polarization or $\chi $-polarization functions used for generating off-diagonal deformations:
\begin{enumerate}
\item[{a]}] The prime metrics inside BHs consist a standard case when the Bekenstein-Hawking thermodynamic paradigm \cite{bek2,haw2} is applicable and
respective hypersurface thermodynamic variables $^{BH}S(A_{Sch}),\ ^{BH}E(A_{Sch})$ and $\ ^{BH}temp(A_{Sch}),$ are computed as stated in \cite%
{hawking73,misner73,wald82} using the area $A_{Sch}$ of a sphere with horizon $\mathring{h},$ see respective formulas (\ref{pm1}) and explanations
for (\ref{metricf}) and (\ref{massf}).

\item[{b]}] A $\tau $-family of d-metrics $\ ^{q}\mathbf{g}(\tau )\rightarrow \ _{Sch}^{int}\mathbf{g}(\tau )=$ $\widehat{\mathbf{g}}(\tau ,\tilde{f}%
(r,\theta ),\tilde{\eta}_{4}(\varphi ))$ describing for $\tau =\tau _{0}$ off-diagonal quasi-stationary configurations (\ref{newbh}), see explanations
related to formulas (\ref{metrfa}), when $h_{4}=\eta _{4}\mathring{g}_{4}=\tilde{f}(r,\theta ),\tilde{\eta}_{4}(\varphi ).$ The gravitational $\eta $%
-polarizations can be such arbitrary ones but also subjected to nonlinear symmetries (\ref{nlims2}), which allows to introduce effective $\Lambda
(\tau ).$ For such data, we can compute the respective volume forms $\ _{\eta }^{J}\mathcal{V[}\ _{Sch}^{int}\mathbf{g}(\tau )]$ (\ref{volumf1})
and use them for defining and computing G. Perelman thermodynamic variables (\ref{thermvar2}). If we prescribe a horizon $\mathring{h}$
determined by a primary Schwarzschild metric (\ref{pm1}), or primary d-metric (\ref{pm1a}), we can formally compute  the Bekenstein-Hawking thermodynamic
variables. Nevertheless, we have to consider additionally the G. Perelman's approach with formulas (\ref{thermvar2}) for $\ _{\eta }^{J}\mathcal{V[}\
_{Sch}^{int}\mathbf{g}(\tau )]$ to characterize more rich off-diagonal configurations encoded in $\tau $-families of solutions (\ref{newbh}).
%%%%%%

\item[{c]}] The geometric flow thermodynamic variables (\ref{thermvar3}) can be computed for a $\tau $-family of quasi-stationary d-metrics 
$\ ^{q}\mathbf{g}(\tau )\rightarrow \ _{Sch}^{int}\mathbf{g}(\chi (\tau ))=\widehat{\mathbf{g}}(\chi _{4}(\tau ,r,\theta ),\mathring{g}_{\alpha \beta })$
defining for $\tau =\tau _{0}$ small $\kappa $-parametric off-diagonal deformations a prime BH d-metric (\ref{pm1a}) with new coefficients being
anisotropic on $\varphi $-coordinate. This refers to $\tau $-running d-metrics of type (\ref{offdnceleps1}). The effective $\tau $-running cosmological constants $\Lambda (\tau )$ can be introduced for nonlinear symmetries (\ref{nlims2}) including $\kappa $-decompositions. For rotoid generating functions (\ref{nlims2}), we can also compute Bekenstein-Hawking thermodynamic variables (they substitute for BE deformations of horizons the similar ones for the Schwarzschild BH). In the G. Perelman approach, we have to compute the respective volume forms $\ _{\chi }^{J}\mathcal{V[}\ _{Sch}^{int}\mathbf{g}(\chi (\tau ))]$ (\ref{volumf1}) and use in formulas for geometric flow thermodynamic variables (\ref{thermvar3}). All discussion after that formulas on certain correlations to $^{BH}S(A_{BE}(\tau )),\ ^{BH}E(A_{BE}(\tau ))$ and $\ ^{BH}temp(A_{BE}(\tau ))$ is valid for this class of off-diagonal quasi-stationary deformations of BH metrics. We have to consider dualization of formulas for a time-like coordinate in interior of a rotoid (or spherical) horizon.
%%%%%%%

\item[{d]}] We can consider gravitational $\chi $-polarizations defining $\tau $-running d-metrics of type (\ref{offdnceleps1}) involving a h-part as
solutions for vortices of generalized Taubes equation (\ref{taubeq}). The effective sources and nonlinear symmetries are correspondingly modified with
such a void $\psi (x^{k})$ function. The corresponding volume functional $\ \ _{\chi }^{J}\mathcal{V[}\ _{Sch}^{int}\mathbf{g}(\psi (void),\chi (\tau
))] $ (\ref{volumf1}) can be computed in $\kappa $-parametric form, which defines respective geometric flow thermodynamic variables (\ref{thermvar3}).

\item[{e]}] In a similar form, we can study gravitational $\chi $-polarizations defining $\tau $-running d-metrics of type (\ref{offdnceleps1}) involving as v-part $\underline{\eta }=\eta _{4}(r,\theta ,\varphi )\rightarrow \underline{\chi }(r,\theta ,\varphi )$ defining solitonic waves (\ref{solitonw}) for a small parameter $\kappa .$ The effective sources and nonlinear symmetries are correspondingly modified for solitonic (radial or angular) configurations for computing respective volume forms $\ _{\chi }^{J}\mathcal{V[}\ _{Sch}^{int}\mathbf{g}(soliton,\underline{\chi }(\tau ))]$ (\ref{volumf1}).
\end{enumerate}
%%%%%%

The above examples a-e] describe a new nonlinear physics encoded in interior of BHs if respective classes of off-diagonal solutions are used for modeling
physical properties of gravitational vacuum under possible relativistic geometric evolution. We emphasize that $\Lambda (\tau )$  are defined because of nonlinear symmetries (\ref{nlims2}) \ of such $\tau $-families of solutions. For certain models, we can consider limits $\Lambda (\tau)\rightarrow 0.$ In this work, we study off-diagonal deformations of the Schwarzschild BHs (for (a)dS configurations, similar constructions were analyzed in \cite{vi17}).
%%%%%%

\subsubsection{Nonholonomic Ricci soliton thermodynamics outside BHs}

In this subsection, we show how thermodynamic variables are computed for prime and target metrics describing regular BH and off-diagonal quasi-stationary deformations as we constructed in section \ref{sec4}.

The prime metrics with mass functions of type $m(r)$ (\ref{massfreg}), $_{e}^{\Lambda }m(r)$ (\ref{massf1}) and $_{e}^{l}m(r)$ (\ref{massf2}) can be
characterized thermodynamically in the framework of the Bekenstein-Hawking paradigm for respective spherical symmetries outside horizons, which can
include also Cauchy or other types horizons. The physical values can be correspondingly labeled by an integer $n,$ effective geometric data $%
(e,\Lambda ),$ and integers $(n,l).$ Similar labels can be used for abstract geometric notations of primary d-metrics, $^{n}\mathbf{\mathring{g},}$ $%
_{e}^{\Lambda }\mathbf{\mathring{g},}$ and $_{l}^{n}\mathbf{\mathring{g}.}$

We can generalize above mentioned prime d-metrics to $\tau $-running off-diagonal quasi-stationary configurations with respective generating functions,  $^{n}\mathbf{g}(\tau )\mathbf{=g}(\Lambda (\tau ),\Phi (\tau ),$ $^{n}\mathbf{\mathring{g}),}\ _{\eta }^{n}\mathbf{g}(\tau )=\ _{\eta }\mathbf{g}(\Lambda (\tau ),$ $^{n}\mathbf{\mathring{g})}$ and $\ _{\chi }^{n}\mathbf{g}(\tau )\mathbf{=}\ _{\chi }\mathbf{g}(\Lambda (\tau ),$
$^{n}\mathbf{\mathring{g});}$ $_{e}^{\Lambda }\mathbf{g}(\tau )=\mathbf{g}%
(\Lambda (\tau ),\Phi (\tau ),$ $_{e}^{\Lambda }\mathbf{\mathring{g}),}\
_{e\eta }^{\Lambda }\mathbf{g}(\tau )=\ _{\eta }\mathbf{g}(\Lambda (\tau ),$ 
$_{e}^{\Lambda }\mathbf{\mathring{g})}$ and $\ \ _{e\chi }^{\Lambda }\mathbf{%
g}(\tau )=\ _{\chi }\mathbf{g}(\Lambda (\tau ),_{e}^{\Lambda }\mathbf{%
\mathring{g});}$ and $_{l}^{n}\mathbf{g}(\tau )\mathbf{=g}(\Lambda (\tau
),\Phi (\tau ),$ $_{l}^{n}\mathbf{\mathring{g}),}_{l\eta }^{n}\mathbf{g}%
(\tau )\mathbf{=}\ _{\eta }\mathbf{g}(\Lambda (\tau ),$ $_{l}^{n}\mathbf{%
\mathring{g})}$ and $_{l\chi }^{n}\mathbf{g}(\tau )\mathbf{=}\ _{\chi }%
\mathbf{g}(\Lambda (\tau ),_{l}^{n}\mathbf{\mathring{g}).}$ Each of such generalization of solutions can be written as $\tau $-families of respective
d-metrics (\ref{offdiagcosmcsh}), (\ref{offdiagpolfr}), or (\ref{offdncelepsilon}). Then, we can characterise such solutions by corresponding thermodynamic variables of type (\ref{thermvar1}), or (\ref{thermvar2}), or (\ref{thermvar3}). The corresponding 9 different type $\tau $-running quasi-stationary configurations are characterized by a proper volume form (\ref{volume}) or (\ref{volumf1}).

For example, we consider the variant $_{l\eta }^{n}\mathbf{g}(\tau ),\,\ $ when the volume form (\ref{volumf1}) is computed: 
\begin{equation*}
\ _{\eta }^{\mathbf{J}}\mathcal{V[}\ _{l\eta }^{n}\mathbf{g}(\tau )]=\int_{\ 
\widehat{\Xi }}\delta \ _{\eta }\mathcal{V}(\ ^{v}\widehat{\mathbf{J}}(\tau
),\ \eta _{\alpha }(\tau ),_{l}^{n}\mathbf{\mathring{g}}).
\end{equation*}%
This value defines the thermodynamic variables of type (\ref{thermvar2}): 
\begin{eqnarray*}
\ _{\eta }^{q}\widehat{Z}(\tau ) &=&\exp \left[ \frac{1}{8\pi ^{2}\tau ^{2}}%
\ \ _{\eta }^{\mathbf{J}}\mathcal{V}[\ \ _{l\eta }^{n}\mathbf{g}(\tau )]%
\right] ,\ _{\eta }^{q}\widehat{\mathcal{E}}\ (\tau )=\ \frac{1-2\tau \
\Lambda (\tau )}{8\pi ^{2}\tau }\ \ \ _{\eta }^{\mathbf{J}}\mathcal{V[}\
_{l\eta }^{n}\mathbf{g}(\tau )], \\
\ \ \ \ _{\eta }^{q}\widehat{S}(\tau ) &=&-\ _{\chi }^{q}\widehat{W}(\tau )=%
\frac{1-\Lambda (\tau )}{4\pi ^{2}\tau ^{2}}\ \ _{\eta }^{\mathbf{J}}%
\mathcal{V[}\ _{l\eta }^{n}\mathbf{g}(\tau )].
\end{eqnarray*}%
The generating and integrating data for $\ \ _{l\eta }^{n}\mathbf{g}(\tau )$ can be such way chosen that they describe outside a Schwarzschild BH
respective $\tau $-families of Taub voids (\ref{taubeq}) or solitonic waves (\ref{solitonw}). In the framework of G. Perelman's thermodynamic paradigm,
we can describe in a self-consistent form the thermodynamic properties of such off-diagonal solutions. For rotoid configurations (\ref{rotoid}),  we
obtain ellipsoidal horizons, which allow us to introduce also the Bekenstein-Hawking alternative description.

All types of quasi-stationary off-diagonal deformations of Schwarzschild BH configurations are described by a similar behaviour under a $\tau $-running
cosmological constant $\Lambda (\tau ).$ For instance, the entropy functional $\ ^{q}\widehat{S}(\tau )\simeq $ $\frac{1-\Lambda (\tau )}{%
4\pi ^{2}\tau ^{2}}$ \ldots\ and the explicit values depend on volume forms which are different for different classes of solutions and chosen generating
and integration data. For instance, such configurations can define an off-diagonal gravitational background determined by the DE distribution with
evolution on effective temperature $\tau .$ Fixing $\tau _{0},$ we obtain thermodynamic models of certain nonholonomic Ricci soliton configurations.
We can calibrate a class of solutions and compute certain entropic/energetic properties using fractions of type (\ref{fractions}) determined by volume
forms.
%%%%%%%

A class of off-diagonal BH deformations can be stable if certain stability conditions are satisfied (see \cite{vi17} and references therein. We can speculate on moving away by some nonlinear waves of respective deformed BH or BE solutions. A quasi-stationary configuration can be "annihilated", i.e. transformed into an  another type of quasi-stationary solution which can be more convenient for a lower entropy $\ _{\eta }^{J}\widehat{S}(\tau ).$ For other types of generating and integrating data, with respective polarization functions,  we can elaborate on  scenarios of creating deformed BHs under $\tau $-evolution.
%%%%%

\section{Conclusions and perspectives}

\label{sec6}This work is on off-diagonal quasi-stationary generalizations of the diagonal solutions \cite{casadio25,ovalle24,casadio24,ovalle21} which were devoted to studies of singular and regular  BH  configurations with finite energy density everywere. Such a research can be related to the challenging  problem of circumventing the singularity theorem in GR \cite{penrose65,penrose69} (see details in \cite{hawking73,misner73,wald82,kramer03}), with certain solutions which implies the existence of certain non-typical states of matter which could manage to modify, even stop, the collapse process. Achieving this without "exotic" matter, maintaining the Schwarzschild horizon (and more general structures of horizons for Kerr-Reisner-Nordstr\"{o}m solutions) consists of  very difficult
conceptual and technical tasks. The problems became more complicated after the  discovery of the late-time cosmic acceleration \cite{riess98,perlmutter99}, which resulted in a plethora of MGTs and scenarios on DE and DM physics   \cite{sotiriou10,nojiri11,capo11,clifton12,harko14,copeland06}. 
%%%%%

\vskip4pt We argue that it is almost impossible to solve about mentioned problems in the framework of GR only using diagonal ansatz, with certain high symmetries (spherical, cylindric, ..., and their small deformations). Such ansatz for metrics transform the (modified) Einstein equations into some systems of nonlinear ODEs which can be more "easy" solved and physically interpreted in terms of certain integration constants. In our approach, we elaborated on new geometric and analytic methods (the so-called AFCDM) \cite{sv11,vacaruplb16,vbubuianu17,partner02}:  It allow us to decouple and solve directly physically important nonlinear systems of PDEs for general off-diagonal metric ansatz with coefficients depending, in principle, on all spacetime coordinates. The AFCDM can be applied for constructing new classes of exact and parametric solutions in GR and various types of MGTs.
%%%%%

\vskip4pt The four-fold aim of this work is stated for constructing and study physical implications of new classes of  solutions in GR describing off-diagonal quasi-stationary deformations of regular primary BH solutions. For simplicity, we consider prime metrics with a Schwarzschild BH horizon and related Cauchy other type horizons as in \cite{casadio25,ovalle24,casadio24}. Four objectives, Obj1 - Obj4 (stated at the end of Introduction, for generating new target solutions) were achieved in such forms:
%%%%%

\vskip4pt In section \ref{sec2} and Appendix \ref{appendixa}, we outlined the main concepts and abstract and coefficient formulas of the AFCDM, which can be used for decoupling and integration in certain general off-diagonal forms the Einstein equations. For such constructions in GR the key idea is to consider (\ref{twocon}) with an auxiliary linear connection, $\widehat{\mathbf{D}}[\mathbf{g}],$ defined for the same metric structure $\mathbf{g}$ as with a (standard) LC-connection $\nabla \lbrack \mathbf{g}].$ We use general quasi-stationary off-diagonal ansatz (with a time-like Killing symmetry) written on Lorentz manifolds in canonical nonholonomic dyadic variables. The resulting system of nonlinear PDEs can be solved exactly or parametrically in general off-diagonal forms. Having defined a general class of quasi-stationary or locally anisotropic solutions, we can always impose additional constraints on generating functions and generating sources which allow us to extract LC configurations. To integrate (modified or not) Einstein equations for a generic off-diagonal ansatz is not possible if we work from the very beginning with $\nabla .$ The main "legal trick" is to distort correspondingly the necessary types of systems of nonlinear PDEs which have decoupling and general integration properties, and then to return to zero nonholonomic torsion (if necessary). In this work, the specific Obj1 was solved in a form describing nonlinear symmetries (relating effective sources to effective cosmological constants) of off-diagonal deformations for regular BH studies in detail in \cite{casadio25}.  
%%%%%

\vskip4pt Our "non-Orthodox" idea is that the singular or regular structure (we do not speculate in this work on scenarios of collapse) is modified by off-diagonal interactions inside the BHs is confirmed by the results of section \ref{sec3}. It is confirmed by the solutions constructed  in the framework of  Obj2. We prove that off-diagonal gravitational interactions modify via gravitational polarizations  the effective physical constants inside the horizons. Corresponding generating functions with gravitational polarizations can be prescribed in general forms, with dependence on a small deformation parameter, admitting void or solitonic like configurations, deformations to rotoid configurations etc. Such a rich internal structure is characterised by specific nonlinear symmetries relating generating functions, generating sources and (effective) cosmological constants.
%%%%%

\vskip4pt To study quasi-stationary off-diagonal deformations of (primary) regular BHs with Schwarzschild exterior studied  in \cite{casadio25,ovalle24} was  stated by Obj3. The physical properties of such new classes of off-diagonal solutions in GR depend positively both on primary data for (diagonal) regular BHs and on the type of gravitational polarizations used for generating target off-diagonal metrics.  The prime configurations are determined by respective mass functions of type $m(r)$ (\ref{massfreg}), $_{e}^{\Lambda }m(r)$ (\ref{massf1}) and $_{e}^{l}m(r)$ (\ref{massf2}) but the generated target solutions can be defined classes of d-metrics (\ref{offdiagcosmcsh}), (\ref{offdiagpolfr}), or (\ref{offdncelepsilon}) with respective generated data. At least for small $\kappa $-parameters, new classes of solutions were constructed as regular ones  with nontrivial effective pressure, non-zero effective cosmological constants,  vanishing angular pressure, etc. In general, such regular BH deformations do not possess horizon, warped, holographic etc. properties and can't be characterized in the framework of Bekenstein-Hawking paradigm for BH solutions \cite{bek2,haw2}.
%%%%%

\vskip4pt  The Obj4,  provides a new thermodynamic paradigm in GR due to relativistic and nonholonomic generalizations of G. Perelman's concept of
W-entropy \cite{perelman1}. As we mentioned in section \ref{sec4}, we do not propose in this work to formulate and prove a certain generalization of the
Thorston-Poincar\'{e} conjecture. Our main goal is to extend in a relativistic form  the thermodynamic part of  G. Perelman genial preprint \cite{perelman1}.  We show that the AFCDM allows us to decouple and solve the corresponding relativistic geometric flow equations (in particular, the nonholonomic Ricci soliton equations including the Einstein equations). We explained how the geometric flow thermodynamic (and possible alternatives of Bekenstein-Hawking thermodynamic) variables can be computed for explicit classes a-e] of deformations in the interior of BHs, as a sub-objective of Obj4. Then (as another sub-objective), we discussed how nonholonomic Ricci flow thermodynamic variables can be computed for $\tau $-families of off-diagonal families of solutions describing the outside of regular BHs. Such diagonal  solutions were studied originally in \cite{casadio25,ovalle24}.   For such models, relativistic extensions of G. Perelman's thermodynamic variables can be computed in certain general forms in terms of effective cosmological constants and using respective volume forms.
%%%%%

\vskip4pt Finally, we conclude that the results and methods of this work can be used for constructing and study in the framework of GR new types of off-diagonal deformations. Such quasi-stationary or locally anisotropc configurations can be  for prime singular or regular  BHs,  describing nonholonomic  Einstein-Yang-Mills-Higgs-Dirac systems, provide various  applications in DE and DM physics, in classical and quantum flow information theories and quantum gravity. Such  partner works were published recently for MGTs, see \cite{partner06,bgrg24,bapny24,vcqg25} and references therein.
%%%%%

\vskip5pt \textbf{Acknowledgement:}\ SV's work is for a visiting fellowship for the Ko\c{c}aeli University in T\"{u}rkiye. It extends former research programs supported by  Fulbright USA-Romania  and CAS LMU at  Munich, Germany,  fellowships during 2022-2024.

\appendix\setcounter{equation}{0} 
\renewcommand{\theequation}
{A.\arabic{equation}} \setcounter{subsection}{0} 
\renewcommand{\thesubsection}
{A.\arabic{subsection}}

\section{Other types of generating functions and gravitational polarizations}

\label{appendixa}

In this Appendix, we re-write the d-metrics $\mathbf{\hat{g}}[\Psi ,\ \
^{v}\Upsilon ]$ \ (\ref{qeltors}) and $\mathbf{\hat{g}}[\Phi ,\ ^{v}\Upsilon
,\Lambda ]$ (\ref{offdiagcosmcsh}) in certain equivalent forms which are
important for generating parametric solutions and study physical properties
of the off-diagonal solutons.

\subsection{A d-metric coefficient used as a generating function}

Taking the partial derivative on $y^{3}$ of $h_{4}$ from (\ref{dmqs}) allows
us to write $h_{4}^{\ast }=-[\Psi ^{2}]^{\ast }/4\ ^{v}\Upsilon .$
Prescribing certain data for $h_{4}(x^{i},y^{3})$ and $\ ^{v}\Upsilon
(x^{i},y^{3})$, we can compute (up to an integration function) a generating
function $\ \Psi $ (using $[\Psi ^{2}]^{\ast }=\int dy^{3}\ ^{v}\Upsilon
h_{4}^{\ast })$ and work with off-diagonal solutions of type (\ref{qeltors}%
). For certain purposes (in equivalent form), we can consider generating
data $(h_{4},\ ^{v}\Upsilon )$ and re-write the quadratic element (\ref%
{qeltors}) in the form 
\begin{eqnarray}
d\widehat{s}^{2} &=&\widehat{g}_{\alpha \beta }(x^{k},y^{3};h_{4},\
^{v}\Upsilon )du^{\alpha }du^{\beta }  \label{offdsolgenfgcosmc} \\
&=&e^{\psi (x^{k})}[(dx^{1})^{2}+(dx^{2})^{2}]-\frac{(h_{4}^{\ast })^{2}}{%
|\int dy^{3}[\ \ ^{v}\Upsilon h_{4}]^{\ast }|\ h_{4}}\{dy^{3}+\frac{\partial
_{i}[\int dy^{3}(\ ^{v}\Upsilon )\ h_{4}^{\ast }]}{\ ^{v}\Upsilon \
h_{4}^{\ast }}dx^{i}\}^{2}  \notag \\
&&+h_{4}\{dt+[\ _{1}n_{k}+\ _{2}n_{k}\int dy^{3}\frac{(h_{4}^{\ast })^{2}}{%
|\int dy^{3}[\ ^{v}\Upsilon h_{4}]^{\ast }|\ (h_{4})^{5/2}}]dx^{k}\}.  \notag
\end{eqnarray}%
So, prescribing $h_{4}(x^{i},y^{3})$ and $\ ^{v}\Upsilon (x^{i},y^{3})$ for (%
\ref{offdsolgenfgcosmc}), we generate quasi-stationary solutions for (\ref%
{cdeq1}).

The nonlinear symmetries (\ref{ntransf1}) and (\ref{ntransf2}) allow to
perform similar computations and relate (\ref{offdsolgenfgcosmc}) to (\ref%
{offdiagcosmcsh}). We have to expres $\Phi ^{2}=-4\ \Lambda h_{4}$ and then
eliminate $\Phi $ from the nonlinear element and generate a solution of (\ref%
{cdeq1a}) beomg type (\ref{offdsolgenfgcosmc}). Such quasi-stationary
off-diagonal metrics are determined by the generating data $(h_{4};\Lambda
,\ ^{v}\Upsilon ).$

\subsection{Quasi-stationary ansatz with gravitational polarizations}

\label{assgravpol}Another type of off-diagonal ansatz allow us to describe
off-diagonal deformations of some prime d-metrics into certain target
d-metrics. We denote a \textit{prime} d-metric as 
\begin{equation}
\mathbf{\mathring{g}=}[\mathring{g}_{\alpha },\mathring{N}_{i}^{a}]
\label{offdiagpm}
\end{equation}%
and transform it into a \textit{target} d-metric $\mathbf{g,}$ 
\begin{equation}
\mathbf{\mathring{g}}\rightarrow \mathbf{g}=[g_{\alpha }=\eta _{\alpha }%
\mathring{g}_{\alpha },N_{i}^{a}=\eta _{i}^{a}\ \mathring{N}_{i}^{a}],
\label{offdiagdefr}
\end{equation}%
which is parameterized as a quasi-stationary d-metric (\ref{dmq}). The
functions $\eta _{\alpha }(x^{k},y^{3})$ and $\eta _{i}^{a}(x^{k},y^{3})$
from (\ref{offdiagdefr}) are called \textit{gravitational polarization (}$%
\eta $\textit{-polarization) functions.} To generate solutions of (\ref%
{cdeq1}) or (\ref{cdeq1a}) we can consider that the nonlinear symmetries (%
\ref{ntransf1}) can be parameterized in certain equivalent forms, 
\begin{eqnarray}
(\Psi ,\ ^{v}\Upsilon ) &\leftrightarrow &(\mathbf{g},\ ^{v}\Upsilon
)\leftrightarrow (\eta _{\alpha }\ \mathring{g}_{\alpha }\sim (\zeta
_{\alpha }(1+\kappa \chi _{\alpha })\mathring{g}_{\alpha },\ ^{v}\Upsilon
)\leftrightarrow  \label{nonlintrsmalp} \\
(\Phi ,\ \Lambda ) &\leftrightarrow &(\mathbf{g},\ \Lambda )\leftrightarrow
(\eta _{\alpha }\ \mathring{g}_{\alpha }\sim (\zeta _{\alpha }(1+\kappa \chi
_{\alpha })\mathring{g}_{\alpha },\ \Lambda ),  \notag
\end{eqnarray}%
where $\kappa $ is a small parameter $0\leq \kappa <1$ with $\zeta _{\alpha
}(x^{k},y^{3})$ and $\chi _{\alpha }(x^{k},y^{3})$ beomg respective $\kappa $%
-polarization functions. The nonholonomic transforms (\ref{nonlintrsmalp})
are assumed to result in a target d-metric $\mathbf{g}$ generating a
solution (\ref{qeltors}) or, equivalently, (\ref{offdiagcosmcsh}). This is
possible if the $\eta $- and/or $\chi $-polarizations are subjected to the
conditions (\ref{ntransf2}) written respectively in the form: 
\begin{eqnarray}
\partial _{3}[\Psi ^{2}] &=&-\int dy^{3}\ ^{v}\Upsilon \partial
_{3}h_{4}\simeq -\int dy^{3}\ ^{v}\Upsilon \partial _{3}(\eta _{4}\ 
\mathring{g}_{4})\simeq -\int dy^{3}\ ^{v}\Upsilon \partial _{3}[\zeta
_{4}(1+\kappa \ \chi _{4})\ \mathring{g}_{4}],  \notag \\
\Phi ^{2} &=&-4\ \Lambda h_{4}\simeq -4\ \Lambda \eta _{4}\mathring{g}%
_{4}\simeq -4\ \Lambda \ \zeta _{4}(1+\kappa \chi _{4})\ \mathring{g}_{4}.
\label{nonlinsymrex}
\end{eqnarray}

Off-diagonal $\eta $-transforms to d-metrics (\ref{offdiagdefr}) can be
re-defined to be generated by $\psi $- and $\eta $-polarizations, 
\begin{equation}
\psi \simeq \psi (\kappa ;x^{k}),\eta _{4}\ \simeq \eta _{4}(x^{k},y^{3}),
\label{etapolgen}
\end{equation}%
in a form equivalent to (\ref{offdsolgenfgcosmc}). In such a case, the
quasi-stationary quadratic element can be written in the form 
\begin{eqnarray}
d\widehat{s}^{2} &=&\widehat{g}_{\alpha \beta }(x^{k},y^{3};\mathring{g}%
_{\alpha };\psi ,\eta _{4};\ \Lambda ,\ ^{v}\Upsilon )du^{\alpha }du^{\beta
}=e^{\psi }[(dx^{1})^{2}+(dx^{2})^{2}]  \label{offdiagpolfr} \\
&&-\frac{[\partial _{3}(\eta _{4}\ \mathring{g}_{4})]^{2}}{|\int dy^{3}\ \
^{v}\Upsilon \partial _{3}(\eta _{4}\ \mathring{g}_{4})|\ \eta _{4}\mathring{%
g}_{4}}\{dy^{3}+\frac{\partial _{i}[\int dy^{3}\ ^{v}\Upsilon \partial
_{3}(\eta _{4}\mathring{g}_{4})]}{\ \ ^{v}\Upsilon \partial _{3}(\eta _{4}%
\mathring{g}_{4})}dx^{i}\}^{2}  \notag \\
&&+\eta _{4}\mathring{g}_{4}\{dt+[\ _{1}n_{k}+\ _{2}n_{k}\int dy^{3}\frac{%
[\partial _{3}(\eta _{4}\mathring{g}_{4})]^{2}}{|\int dy^{3}\ \ ^{v}\Upsilon
\partial _{3}(\eta _{4}\mathring{g}_{4})|\ (\eta _{4}\mathring{g}_{4})^{5/2}}%
]dx^{k}\}^{2}.  \notag
\end{eqnarray}%
This ansatz can be related to a solution of type (\ref{offdiagcosmcsh}) if $%
\Phi ^{2}=-4\ \Lambda h_{4}$ and the $\eta $-polarizations are determined by
the generating data $(h_{4}=\eta _{4}\mathring{g}_{4};\Lambda ,\
^{v}\Upsilon ).$

If a primary d-metric $\mathbf{\mathring{g}}$ (\ref{offdiagpm}) is chosen,
for instance, as a regular BH solution from \cite{casadio25}, we can analyze
possible embedding into a quasi-stationary background determined by a target 
$\mathbf{g}$ (\ref{offdiagdefr}). In general, an off-diagonal solution (\ref%
{offdiagpolfr}) do not define a singular or regular BH. We need additional
assumptions and a rigorous analysis to distinguish certain cases when a
target d-metric has 1) an unclear physical meaning; or 2) it describes a BH
embedded in a nontrivial gravitational vacuum determined by $\eta $%
-polarizations, for instance, taken as certain solitonic waves, and/or 3)
related to pattern forming structures, spacetime quasi-cristals etc.; or 4)
we generate a regular/ singular BH solution with possible small $\chi $%
-polarizations of physical constants and horizons. An additional analysis
similar to that in \cite{vbubuianu17} (and references therein) is necessary.
Here, we also note that non-stable solutions also have physical importance
in physics. They may describe 5) phase transitions of different types of
BHs, explosions of BHs etc. which are modelled by nonlinear off-diagonal
interactions. In all cases of solutions of type (\ref{offdiagcosmcsh}) or (%
\ref{offdiagpolfr}), we can compute generalized G. Perelman variables \cite%
{perelman1,svnonh08,gheorghiuap16,partner06}, see section \ref{sec5}. If (%
\ref{offdiagpolfr}) involves some horizons, we can consider the
Bekenstein-Hawking thermodynamic paradigm \cite{bek2,haw2}.

\subsection{Small parametric off-diagonal deformations}

We can introduce $\kappa $-linear nonlinear transforms (\ref{nonlintrsmalp})
with generating functions involving $\chi $-polarizations in (\ref%
{offdiagpolfr}). This defines small nonholonomic deformations of a prime
d-metric $\mathbf{\mathring{g}}$ into so-called $\kappa $-parametric
solutions with $\zeta $- and $\chi $-coefficients when 
\begin{eqnarray*}
\psi &\simeq &\psi (x^{k})\simeq \psi _{0}(x^{k})(1+\kappa \ _{\psi }\chi
(x^{k})),\mbox{ for }\  \\
\ \eta _{2} &\simeq &\eta _{2}(x^{k_{1}})\simeq \zeta _{2}(x^{k})(1+\kappa
\chi _{2}(x^{k})),\mbox{ we can consider }\ \eta _{2}=\ \eta _{1}; \\
\eta _{4} &\simeq &\eta _{4}(x^{k},y^{3})\simeq \zeta
_{4}(x^{k},y^{3})(1+\kappa \chi _{4}(x^{k},y^{3})).
\end{eqnarray*}%
In these formulas, $\psi $ and $\eta _{2}=\ \eta _{1}$ can be chosen to be
related to the solutions of the 2-d Poisson equation $\partial _{11}^{2}\psi
+\partial _{22}^{2}\psi =2\ ^{v}\Upsilon (x^{k}).$ As a result, we compute $%
\kappa $-parametric deformations to quasi-stationary d-metrics with $\chi $%
-generating functions, 
\begin{equation*}
d\ \widehat{s}^{2}=\widehat{g}_{\alpha \beta }(x^{k},y^{3};\psi
,g_{4};^{v}\Upsilon )du^{\alpha }du^{\beta }=e^{\psi _{0}}(1+\kappa \ ^{\psi
}\chi )[(dx^{1})^{2}+(dx^{2})^{2}]
\end{equation*}%
\begin{eqnarray*}
&&-\{\frac{4[\partial _{3}(|\zeta _{4}\mathring{g}_{4}|^{1/2})]^{2}}{%
\mathring{g}_{3}|\int dy^{3}\{\ \ ^{v}\Upsilon \partial _{3}(\zeta _{4}%
\mathring{g}_{4})\}|}-\kappa \lbrack \frac{\partial _{3}(\chi _{4}|\zeta _{4}%
\mathring{g}_{4}|^{1/2})}{4\partial _{3}(|\zeta _{4}\mathring{g}_{4}|^{1/2})}%
-\frac{\int dy^{3}\{\ ^{v}\Upsilon \partial _{3}[(\zeta _{4}\mathring{g}%
_{4})\chi _{4}]\}}{\int dy^{3}\{\ ^{v}\Upsilon \partial _{3}(\zeta _{4}%
\mathring{g}_{4})\}}]\}\mathring{g}_{3} \\
&&\{dy^{3}+[\frac{\partial _{i}\ \int dy^{3}\ ^{v}\Upsilon \ \partial
_{3}\zeta _{4}}{(\mathring{N}_{i}^{3})\ ^{v}\Upsilon \partial _{3}\zeta _{4}}%
+\kappa (\frac{\partial _{i}[\int dy^{3}\ ^{v}\Upsilon \ \partial _{3}(\zeta
_{4}\chi _{4})]}{\partial _{i}\ [\int dy^{3}\ ^{v}\Upsilon \partial
_{3}\zeta _{4}]}-\frac{\partial _{3}(\zeta _{4}\chi _{4})}{\partial
_{3}\zeta _{4}})]\mathring{N}_{i}^{3}dx^{i}\}^{2}
\end{eqnarray*}%
\begin{eqnarray}
&&+\zeta _{4}(1+\kappa \ \chi _{4})\ \mathring{g}_{4}\{dt+[(\mathring{N}%
_{k}^{4})^{-1}[\ _{1}n_{k}+16\ _{2}n_{k}[\int dy^{3}\frac{\left( \partial
_{3}[(\zeta _{4}\mathring{g}_{4})^{-1/4}]\right) ^{2}}{|\int dy^{3}\partial
_{3}[\ ^{v}\Upsilon (\zeta _{4}\mathring{g}_{4})]|}]  \label{offdncelepsilon}
\\
&&+\kappa \frac{16\ _{2}n_{k}\int dy^{3}\frac{\left( \partial _{3}[(\zeta
_{4}\mathring{g}_{4})^{-1/4}]\right) ^{2}}{|\int dy^{3}\partial _{3}[\
^{v}\Upsilon (\zeta _{4}\mathring{g}_{4})]|}(\frac{\partial _{3}[(\zeta _{4}%
\mathring{g}_{4})^{-1/4}\chi _{4})]}{2\partial _{3}[(\zeta _{4}\mathring{g}%
_{4})^{-1/4}]}+\frac{\int dy^{3}\partial _{3}[\ ^{v}\Upsilon (\zeta _{4}\chi
_{4}\mathring{g}_{4})]}{\int dy^{3}\partial _{3}[\ ^{v}\Upsilon (\zeta _{4}%
\mathring{g}_{4})]})}{\ _{1}n_{k}+16\ _{2}n_{k}[\int dy^{3}\frac{\left(
\partial _{3}[(\zeta _{4}\mathring{g}_{4})^{-1/4}]\right) ^{2}}{|\int
dy^{3}\partial _{3}[\ ^{v}\Upsilon (\zeta _{4}\mathring{g}_{4})]|}]}]%
\mathring{N}_{k}^{4}dx^{k}\}^{2}.  \notag
\end{eqnarray}%
The $\kappa $-parametric \ off-diagonal solutions (\ref{offdncelepsilon})
allow us to define, for instance, ellipsoidal deformations of regular BH
metrics into regular BE ones. For other classes of solutions with small
parameters, we can compute quasi-classical off-diagonal deformations of some
solutions in GR. Various corrections from MGTs (defined by nonassociative/
noncommutative or supersymmetric terms) can be also defined and computed as
certain off-dagonal $\kappa $-parametric solutions.

%\appendix

\setcounter{equation}{0} \renewcommand{\theequation}
{B.\arabic{equation}} \setcounter{subsection}{0} 
\renewcommand{\thesubsection}
{B.\arabic{subsection}}

\section{Relativistic geometric flows of nonholonomic Einstein systems}

\label{appendixb}

In this Appendix, we outline the main results on relativistic geometric flow
evolution of\textit{\ nonholonomic Einstein systems} (NESs) described by $%
\tau $-families of gravitational filed equations written in canonical dyadic
variables $[\mathbf{g}(\tau ),\mathbf{N}(\tau ),\widehat{\mathbf{D}}(\tau )]$
as we stated at the beginning of section \ref{sec5}. Such nonlinear systems
of PDE can be also decoupled and integrated in general off-diagonal forms by
applying the AFCDM. %%%%%

\subsection{Canonical nonholonomic F- and W-functionals}

For elaborating various types of geometric flow theories and applications in
modern physics, the concepts of $\mathcal{F}$- and $\mathcal{W}$-functionals
is very important \cite{perelman1}. We use corresponding relativistic
formulations in canonical dyadic variables \cite%
{svnonh08,gheorghiuap16,partner06} from which generally integrable
nonholonomic geometric flow equations can be proved in variational N-adapted
forms.

We postulate such modified G. Perelman's functionals for describing
relativistic geometric flows of NESs: 
\begin{eqnarray}
\widehat{\mathcal{F}}(\tau ) &=&\int_{t_{1}}^{t_{2}}\int_{\Xi _{t}}e^{-%
\widehat{\zeta }(\tau )}\sqrt{|\mathbf{g}(\tau )|}\delta ^{4}u[\widehat{%
\mathbf{R}}sc(\tau )+\ \widehat{\mathcal{L}}(\tau )+|\widehat{\mathbf{D}}%
(\tau )\widehat{\zeta }(\tau )|^{2}],  \label{fperelmNES} \\
\ \widehat{\mathcal{W}}(\tau ) &=&\int_{t_{1}}^{t_{2}}\int_{\Xi _{t}}\left(
4\pi \tau \right) ^{-2}e^{-\widehat{\zeta }(\tau )}\sqrt{|\mathbf{g}(\tau )|}%
\delta ^{4}u[\tau (\widehat{\mathbf{R}}sc(\tau )+\widehat{\mathcal{L}}(\tau
)+|\widehat{\mathbf{D}}(\tau )\widehat{\zeta }(\tau )|^{2})+\widehat{\zeta }%
(\tau )-4].  \label{wfperelmNES}
\end{eqnarray}%
In these formulas, we consider $\tau $-families of d-objects (\ref{twocon}),
(\ref{criccidt}) and (\ref{criccidsc}). The term $\widehat{\mathcal{L}}(%
\widehat{\mathbf{R}}sc(\tau ))=\widehat{\mathbf{R}}sc(\tau )$ can be
considered for $\ $%
\begin{equation*}
\mathcal{L}(Rsc(g(\tau ),\nabla (\tau ))=\ \widehat{\mathcal{L}}(\widehat{%
\mathbf{R}}sc(\tau ))+\ ^{e}\widehat{\mathcal{L}}(g(\tau ),\widehat{\mathbf{Z%
}}(\tau ))
\end{equation*}%
derived by using canonical distortion relations (\ref{canondistrel}). The
effective Lagrange densities $\ \widehat{\mathcal{L}}(\tau )=\ ^{e}\ 
\widehat{\mathcal{L}}(\tau )+\ \ ^{m}\widehat{\mathcal{L}}(\tau )$ are
introduced as we explain for (\ref{emdt}) and (\ref{einstdist}). Above
integrals are defined and computed for a 3+1 splitting of $\tau $-familes of
geometric objects on a nonholonomic Lorentz manifold $\mathbf{V}$, for some
4-d regions defined by space like fibrations $(t,\Xi _{t};t\in \lbrack
t_{1}<t_{2}])$ \cite{misner73}. This is necessary for elaborating
thermodynamic models. A spacetime $\mathbf{V}$ involves also a nonholonomic
2+2 splitting which is necessary for generating off-diagonal solutions.

To elaborate geometric and physical models certain normalization conditions,%
\begin{equation}
\int_{t_{1}}^{t_{2}}\int_{\Xi _{t}}\left( 4\pi \tau \right) ^{-2}e^{-%
\widehat{\zeta }(\tau )}\sqrt{|\mathbf{g}|}d^{4}u=1,  \label{normcond}
\end{equation}%
are imposed, where $\widehat{\zeta }(\tau )=\widehat{\zeta }(\tau ,u)$ is a
family normalizing functions. In some recent works related to string theory 
\cite{papad24}, $\widehat{\zeta }(\tau )$ are treated as $\tau $-dilatonic
fields but this depends on the type of Ricci flow model we elaborate \cite%
{partner06}. In our wors, we assume that we can chose any type of
normalization which simplify certain derived systems of nonlinear PDEs and
allow to compute certain thermodynamic variables in explicit forms. We can
extract geometric flows of LC configurations if we consider $\tau $-families
of additional nonholonomic constraints (\ref{lccond1}) or (\ref{lccond}).
The normalizing function, $\widehat{\zeta }(\tau )\rightarrow $ $\zeta (\tau
),$ from (\ref{fperelmNES}) and (\ref{wfperelmNES}) can be re-defined in
other different forms which allow to generate, or absorb, respective
distortions of connections and work with respective functionals for other
geometric data. For instance, we can consider $\mathcal{F}(\tau ,\zeta ,%
\mathbf{g},\mathbf{D},\mathbf{R}sc,...)$ and $\mathcal{W}(\tau ,\zeta ,%
\mathbf{g},\mathbf{D},\mathbf{R}sc,...)$ defined by using an arbitrary
metric-affine d-connection.

\subsection{Geometric flow equations and nonholonomic Ricci solitons as
Einstein spaces}

In \cite{svnonh08,gheorghiuap16,partner06}, we emphasize that there are two
possibilities to derive geometric flow equations from a functional $\widehat{%
\mathcal{F}}(\tau )$ or $\widehat{\mathcal{W}}(\tau ).$ In the first case,
we can use $\widehat{\mathbf{D}}(\tau )$ instead of $\nabla (\tau )$ and
reproduce in N-adapted and distorted forms all covariant differential and
integral formulas considered in \cite{perelman1,monogrrf1,monogrrf2,monogrrf3}. Such rigorous proofs can be
written on hundred of pages and involve many problems for extension to
various types of MGT if certain nonassociative/ noncommutative or
supersymmetric variables are involved.  In the second case, we can use an
abstract geometric formalism \cite{misner73,svnonh08,gheorghiuap16} by
involving necessary types of  nonholonomic distortions and dyadic variables.

In hat variables, we postulate such relativistic generalizations of the
Hamilton-Friedan equations \cite{friedan80,hamilton82}: 
\begin{eqnarray}
\partial _{\tau }g_{ij}(\tau ) &=&-2[\widehat{\mathbf{R}}_{ij}(\tau
)-\Upsilon _{ij}(\tau )];\ \partial _{\tau }g_{ab}(\tau )=-2[\widehat{%
\mathbf{R}}_{ab}(\tau )-\ \Upsilon _{ab}(\tau )];  \label{ricciflowr2} \\
\widehat{\mathbf{R}}_{ia}(\tau ) &=&\widehat{\mathbf{R}}_{ai}(\tau )=0;%
\widehat{\mathbf{R}}_{ij}(\tau )=\widehat{\mathbf{R}}_{ji}(\tau );\widehat{%
\mathbf{R}}_{ab}(\tau )=\widehat{\mathbf{R}}_{ba}(\tau );\   \notag \\
\partial _{\tau }\widehat{\zeta }(\tau ) &=&-\widehat{\square }(\tau )[%
\widehat{\zeta }(\tau )]+\left\vert \widehat{\mathbf{D}}(\tau )[\widehat{%
\zeta }(\tau )]\right\vert ^{2}-\ \widehat{R}sc(\tau )+\ \Upsilon _{\ \alpha
}^{\alpha }(\tau ).  \notag
\end{eqnarray}%
In these formulas, a $\tau $-family of effective sources as for (\ref{esourc}%
), $\widehat{\mathbf{\Upsilon }}_{\ \ \beta }^{\alpha }(\tau )=[\
^{h}\Upsilon (\tau )\delta _{\ \ j}^{i},\ ^{v}\Upsilon (\tau )\delta _{\ \
b}^{a}]$ is used; and a generalized Laplace operator $\widehat{\square }%
(\tau )=\widehat{\mathbf{D}}^{\alpha }(\tau )\widehat{\mathbf{D}}_{\alpha
}(\tau )$ is used. Here we note that the conditions $\widehat{\mathbf{R}}%
_{ia}=\widehat{\mathbf{R}}_{ai}=0$ for the Ricci tensor $\widehat{R}ic[%
\widehat{\mathbf{D}}]=\{\widehat{\mathbf{R}}_{\alpha \beta }=[\widehat{R}%
_{ij},\widehat{R}_{ia},\widehat{R}_{ai},\widehat{R}_{ab,}]\}$ have to be
imposed in (\ref{ricciflowr2}) if we want to keep the metrics $\mathbf{g}%
(\tau )$ to be symmetric. In general, nonholonomic and nonmetric Ricci flow
evolution scenarios can be elaborated if such conditions are dropped, see a
review in \cite{partner02,partner06} and references therein.

We can reproduce all solutions from the sections \ref{sec3} and \ref{sec4}
by introducing a formal dependence on and re-defining the effective sources
in the form 
\begin{eqnarray}
\widehat{\mathbf{J}}(\tau ) &=&\widehat{\mathbf{\Upsilon }}(\tau )-\frac{1}{2%
}\partial _{\tau }\mathbf{g}(\tau )=[\ ^{h}\widehat{\mathbf{J}}(\tau ),\ ^{v}%
\widehat{\mathbf{J}}(\tau )]  \label{effrfs} \\
&=&[\ J_{i}(\tau )=\ \Upsilon _{i}(\tau )-\frac{1}{2}\partial _{\tau
}g_{i}(\tau ),\ \ J_{a}(\tau )=\Upsilon _{a}(\tau )-\frac{1}{2}\partial
_{\tau }g_{a}(\tau )],  \notag
\end{eqnarray}%
where $\mathbf{g}(\tau )=[g_{i}(\tau ),g_{a}(\tau ),N_{i}^{a}(\tau )]$ are
defined as for (\ref{dm}) and $\widehat{\mathbf{\Upsilon }}_{\alpha }(\tau
)=[\Upsilon _{i}(\tau ),\Upsilon _{a}(\tau )]$ for (\ref{esourc}). For such
effective sources $\widehat{\mathbf{J}}(\tau )$ (\ref{effrfs}), the
nonholonomic variables can be adapted in such forms that the off-diagonal
solutions of (\ref{cdeq1}) are transformed into $\tau $-depending solutions
of (\ref{ricciflowr2}). In respective formulas, we change $\widehat{\mathbf{%
\Upsilon }}_{\alpha }\rightarrow \widehat{\mathbf{J}}(\tau )$ and, for
instance, $\tau $-families of quasi-stationary d-metrics (\ref{dmq}), can be
generated for respective ansatz with $g_{i}(\tau )=e^{\psi {(\tau ,x^{j})}},$
$g_{a}(\tau )=h_{a}(\tau ,x^{k},y^{3}),$ $\ N_{i}^{3}=w_{i}(\tau
,x^{k},y^{3})$ and $\,\,\,\,N_{i}^{4}=n_{i}(\tau ,x^{k},y^{3}).$

Nonholonomic Ricci solitons were defined and studied in \cite%
{svnonh08,gheorghiuap16} as self-similar configurations for $\tau =\tau _{0}$
in the corresponding nonholonomic geometric flow equations. Using (\ref%
{ricciflowr2}), we obtain the equations for the nonholonomic Einstein
equations with effective sources (\ref{effrfs}) and (\ref{esourc}), 
\begin{align}
\widehat{\mathbf{R}}_{ij}& =\ ^{h}\widehat{\mathbf{J}}(\tau _{0},{x}^{k}),\ 
\widehat{\mathbf{R}}_{ab}=\ ^{v}\widehat{\mathbf{J}}(\tau _{0},x^{k},y^{c}),
\label{canriccisolda} \\
\widehat{\mathbf{R}}_{ia}& =\widehat{\mathbf{R}}_{ai}=0;\widehat{\mathbf{R}}%
_{ij}=\widehat{\mathbf{R}}_{ji};\widehat{\mathbf{R}}_{ab}=\widehat{\mathbf{R}%
}_{ba}.  \notag
\end{align}%
For additional constraints (\ref{lccond1}) or (\ref{lccond}) for zero
nonholonomic torsion, such equations transform into the standard Einstein
equations $\nabla (\tau _{0}).$ %%%%%%%

\end{document}